\documentclass[journal=jctcce,manuscript=article]{achemso}

\usepackage[version=3]{mhchem} 
\usepackage{bm}
\usepackage{xcolor}
\usepackage{dsfont}
\usepackage[left=2cm,right=2cm,top=3cm,bottom=2cm]{pdflscape}
\usepackage{subcaption}
\usepackage{amssymb}

\author{Joseph P. Heindel}
\affiliation{Kenneth S. Pitzer Theory Center and Department of Chemistry, University of California, Berkeley, California 94720, United States}
\alsoaffiliation{Chemical Sciences Division, Lawrence Berkeley National Laboratory, Berkeley, California 94720, United States}
\author{Lukas Kim}
\affiliation{Kenneth S. Pitzer Theory Center and Department of Chemistry, University of California, Berkeley, California 94720, United States}
\author{Martin Head-Gordon}
\affiliation{Kenneth S. Pitzer Theory Center and Department of Chemistry, University of California, Berkeley, California 94720, United States}
\alsoaffiliation{Chemical Sciences Division, Lawrence Berkeley National Laboratory, Berkeley, California 94720, United States}

\author{Teresa Head-Gordon}
\email{heindelj@lbl.gov, thg@berkeley.edu}
\affiliation{Kenneth S. Pitzer Theory Center and Department of Chemistry, University of California, Berkeley, California 94720, United States}
\alsoaffiliation{Chemical Sciences Division, Lawrence Berkeley National Laboratory, Berkeley, California 94720, United States}
\alsoaffiliation{Departments of Bioengineering and Chemical and Biomolecular Engineering, University of California, Berkeley, California 94720, United States}

\title[An \textsf{achemso} demo]
  {Completely Multipolar Model for Many-Body Water-Ion and Ion-Ion Interactions}


\begin{document}

\begin{tocentry}
\centering
\includegraphics[width=\textwidth]{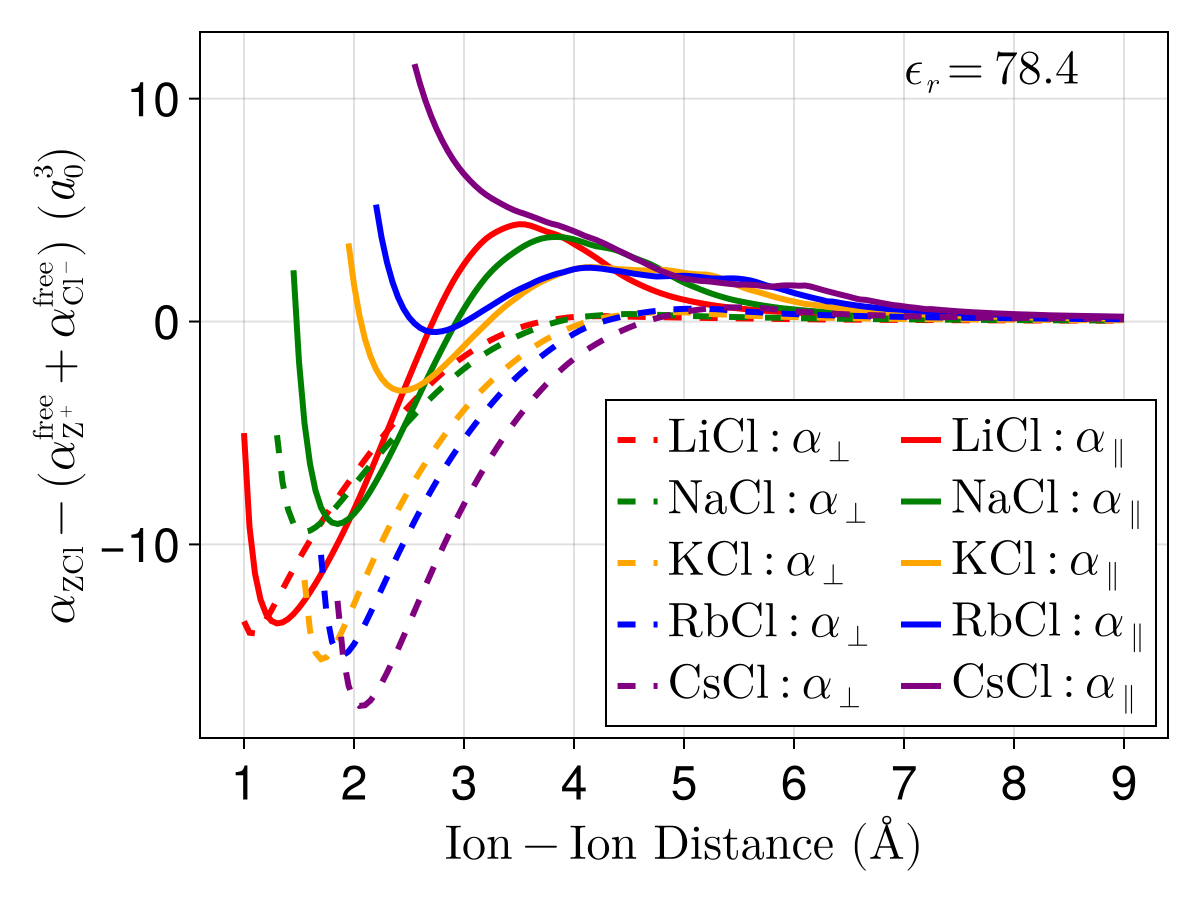}
\end{tocentry}

\begin{abstract}
\noindent
This work constructs an advanced force field, the Completely Multipolar Model (CMM), to quantitatively reproduce each term of an energy decomposition analysis (EDA) for aqueous solvated alkali metal cations and halide anions and their ion pairings. We find that all individual EDA terms remain well-approximated in the CMM for ion-water and ion-ion interactions, except for polarization,  which shows errors due to the partial covalency of ion interactions near their equilibrium. We quantify the onset of the dative bonding regime by examining the change in molecular polarizability and Mayer bond indices as a function of distance, showing that partial covalency manifests by breaking the symmetry of atomic polarizabilities while strongly damping them the at short-range. This motivates an environment-dependent atomic polarizability parameter that depends on the strength of the local electric field experienced by the ions to account for strong damping, with anisotropy introduced by atomic multipoles. The resulting CMM model for ions provides accurate dimer surfaces and three-body polarization and charge transfer compared to EDA, and shows excellent performance on various ion benchmarks including vibrational frequencies and cluster geometries.
\end{abstract}

\section*{Introduction}
Solvated atomic ions such as alkaline metals and halides play critical roles in the regulation of biological systems, environmental chemistry, and energy materials, and hence they are of great interest for molecular simulation. And yet our ability to treat the interactions between water and ions and ion pairs using classical fixed charge force fields (FFs) suffers from large inaccuracies, an open problem which has remained elusive dating back decades.\cite{Jungwirth2006,Mamatkulov2018} As is known, non-additive interactions are critical for describing hydrogen-bonded interactions in water\cite{xantheas2000cooperativity}, and the many-body energy contributions for water interacting with ions is quite large, typically around 15\% of the non-bonded energy, and the cooperativity effects can be either stabilizing or destabilizing depending on ion type and binding motif.\cite{heindel2021many,herman2021many} And yet many simulation models involving ions or ionized molecules are carried out by treating the
ion as a scaled point charge combined with simple functional forms for dispersion and repulsion that cannot describe the inherent many-body complexity.\cite{Joung2008, leontyev2011accounting,bedrov2019molecular} 

We have recently formulated a principled approach to FF development,\cite{mao2017energy,demerdash2017assessing} manifested in the Completely Multipolar Model (CMM)\cite{heindel2024} which we illustrated for water interactions. The CMM model utilizes a general functional form for the damping functions that combines a Slater density with electric multipoles to describe the energy decompositions of the quantum mechanical non-bonded interaction energy using ALMO-EDA\cite{khaliullin2007,horn2016probing,Mao:2021:EDA-review}
\begin{equation}
E_{\rm int} = E_{\rm Elec} + E_{\rm Pauli} + E_{\rm Disp} + E_{\rm Pol} +  E_{\rm CT}
 \label{eq:eda}
\end{equation}
where $E_{\rm Elec}$, $E_{\rm Pauli}$, $ E_{\rm Disp}$, $E_{\rm Pol}$, and  $E_{\rm CT}$ correspond to the contributions from the permanent electrostatics, Pauli repulsion, dispersion, polarization, and charge transfer, respectively. We note that the CMM model also contains a term for the exchange polarization energy that is part of $E_{\rm Pol}$ in Eq. \ref{eq:eda}, which  describes the relief of Pauli repulsion due to polarization.\cite{horn2016probing} We find good agreement with the EDA breakdown term-by-term for water dimers and trimers that leads to excellent reproduction of large water cluster benchmarks and spectroscopic observables.\cite{heindel2024}

This work extends the CMM to the regime of very strong molecular interactions represented by ions. We find that the CMM used for water offers the same excellent results for ion-water and ion-ion interactions when fit against all terms of the EDA in Eq. \ref{eq:eda}, except for polarization which manifests significant errors. In part the systematic error is due to the neglect of higher order terms in polarization, but is further exacerbated by the effects of partial covalency or dative bonding near the equilibrium distance of ion-water and ion pair interactions. The nature of partial covalency is analyzed with recent metrics for chemical bonding, the polarizability\cite{hait2023bond} and bond flux\cite{Kim2024}, that makes clear that the polarization model in CMM must be adapted to describe an environment-dependent modulation of the ion polarizability parameters. With this physically motivated modification, we show that the CMM model conforms well to all EDA terms for ion-water and ion-ion interactions, that in turn yields excellent agreement with independent ion benchmark data including harmonic vibrations and cluster geometries.\cite{boyer2019beyond}

\section*{Theory and Methods}
\noindent
The CMM utilizes multipolar electrical moments modulated by exponential decay of electron density as a common functional form for all terms in Eq. \ref{eq:eda} which we briefly review here. In particular, the Coulomb potential is expanded in gradients of the damped potential which result in the Coulomb interaction tensors for the one-center case, $\bm{T}_{ij}^{damp}$,
\begin{equation}
    \bm{T}_{ij}^{damp}=
    \begin{bmatrix}
        1 & \nabla & \nabla^2
    \end{bmatrix}\left(\frac{f_{ij}^{damp}(r_{ij})}{r_{ij}}\right)
    \label{eq:tensor_1}
\end{equation}
\noindent
and the two-center overlap interaction tensors, $\bm{T}_{ij}^{overlap}$,
\begin{equation}
    \bm{T}_{ij}^{overlap}=
    \begin{bmatrix}
        1 & \nabla & \nabla^2 \\
        \nabla & \nabla^2 & \nabla^3 \\
        \nabla^2 & \nabla^3 & \nabla^4 \\
    \end{bmatrix}\left(\frac{f_{ij}^{overlap}(r_{ij})}{r_{ij}}\right)
    \label{eq:tensor_2}
\end{equation}
\noindent
where the explicit form of the damping and overlap functions for the one-center and two-center cases are\cite{rackers2021polarizable},
\begin{subequations}
    \begin{equation}
        f^{damp}_i(r_{ij})=1-\left(1+\frac{1}{2}(b_ir_{ij})\right)e^{-b_ir_{ij}}
        \label{eq:f_damp}
    \end{equation}
    \begin{equation}
        f^{overlap}_{ij}(r_{ij})=1-\left(1+\frac{11}{16}(b_{ij}r_{ij})+\frac{3}{16}(b_{ij}r_{ij})^2+\frac{1}{48}(b_{ij}r_{ij})^3\right)e^{-b_{ij}r_{ij}}
        \label{eq:f_overlap}
    \end{equation}
\end{subequations}
\noindent
In the CMM approach, Eqs.  \ref{eq:tensor_1} and \ref{eq:tensor_2} generate the damped interaction tensors used for evaluating permanent electrostatics (Elec) and polarization (Pol), and the overlap tensors are used in the Pauli repulsion (Pauli), dispersion (Disp), charge transfer (CT), and exchange polarization (Exch-Pol) energy as well. These damped interaction tensors have the same form as undamped Coulomb interaction tensors except that various entries in the tensor are scaled. Furthermore for Pauli, CT, and Exch-Pol, we use the difference between the damped and undamped multipolar interaction tensors to maintain the short-ranged character of these terms. 
\begin{subequations}
\begin{equation}
    \mathbf{T}^{sr,+}_{ij}=(\mathbf{T}_{ij}-\mathbf{T}_{ij}^{overlap})
    \label{eq:mult_sr_plus}
\end{equation}
\begin{equation}
    \mathbf{T}^{sr,-}_{ij}=(\mathbf{T}_{ij}^{overlap}-\mathbf{T}_{ij})
    \label{eq:mult_sr_minus}
\end{equation}
\end{subequations}
Eq. \ref{eq:mult_sr_plus} is used for Pauli, while Eq. \ref{eq:mult_sr_minus} is used for CT and Exch-Pol. Finally, we allow each short-range term of the EDA in the CMM model to have a different value of $b$, a physical motivation related to the observation that coupling between exchange and each type of interaction occurs at different spatial ranges.\cite{tang1992damping} 

\textbf{Electrostatics and charge penetration}. The CMM uses a traditional point multipole approach up to the quadrupoles for water, and a charge penetration (CP) contribution that modifies the short-range electrostatic energy to be more attractive than the point multipole expansion alone. The  electrostatic energy expression used here for ions is the same as that of water, although it is truncated at the monopole contribution:
\begin{equation}
  V_{elec}=\sum_{i<j}Z_iT_{ij}Z_j+Z_i\bm{T}_{ij}^{damp}\bm{M}_j+Z_j\bm{T}_{ji}^{damp}\bm{M}_i+\bm{M}_i\bm{T}_{ij}^{overlap}\bm{M}_j
  \label{eq:elec}
\end{equation}
where the first term in Eq. \ref{eq:elec} represents repulsive core-core interactions where $T_{ij}=1/r_{ij}$ with $Z_i$ the core charge on the $i$th atom. The second and third terms describe attractive core-shell interactions where $\bm{M}_i$ is a vector whose entries are the components of the multipoles located on that atom. The final term corresponds to the shell-shell interactions. The core-shell and shell-shell interaction tensors are defined in Eqs. \ref{eq:tensor_1} and \ref{eq:tensor_2}.

\textbf{Dispersion}. The dispersion energy uses a damped polynomial interaction given by,
\begin{equation}
  V_{disp}=\sum_{i<j}f_6^{TT}(r_{ij})\frac{C_{6,ij}}{r_{ij}^6}
  \label{eq:disp}
\end{equation}
\noindent
where $C_{6,ij}$ is the dispersion coefficient between atoms $i$ and $j$ which is determined as $C_{6,ij}=\sqrt{C_{6,i}C_{6,j}}$, and $C_{6,i}$ is a parameter fit to the EDA dispersion energy, and $f_6^{TT}(r_{ij})$ is the sixth-order Tang-Toennies (TT) damping function\cite{tang1984improved} 
\begin{equation}
  f_6^{TT}(r_{ij}) = 1-e^{-r_{ij}}\sum_{k=0}^6\frac{r_{ij}^k}{k!}
  \label{eq:TT}
\end{equation}
The dispersion energy expression developed for water is also the same for the case of ions, except that we do not use mixing rules to model the $C_{6,ij}$ but fit the coefficients directly for ion-ion interactions.

\textbf{Pauli Repulsion.} We define the Pauli repulsion energy as,
\begin{equation}
    V_{Pauli}=\sum_{i<j}\mathbf{M}_i^{Pauli}\mathbf{T}_{ij}^{sr,+}\mathbf{M}_j^{Pauli}
    \label{eq:pauli}
\end{equation}
\noindent
where $\mathbf{M}_i^{Pauli}$=$\mathbf{K}_i\mathbf{M}_i^{Elec}$, is a vector representing all multipoles up to quadrupoles for water and monpoles for ions. We explicitly fit only the Pauli charges, $K_{q}$, while forcing the Pauli dipoles, $K_{\mathbf{\mu}}$, and quadrupoles, $K_{\mathbf{\Theta}}$, to be proportional to the electric dipoles and quadrupoles. For water, we make the Pauli repulsion charges, $K_i^q$, dependent on the \ce{O-H} bond length:
\begin{equation}
K_i^q(R_{ij})=K_i^q+j_{b,pauli}^{\mathrm{OH}}(R_{ij}-R_e)
  \label{eq:variable_pauli}
\end{equation}
a motivation described in the CMM paper.\cite{heindel2024}

\textbf{Polarization}. The CMM uses polarizable dipoles\cite{Demerdash2014,Demerdash2018} for ions and water, and incorporates charge flow polarization using a modification of the electronegativity equalization model (EEM) for molecules.\cite{mortier1986electronegativity} The energy of an induced dipole $\bm{\mu}_i^{ind}$ in an electric field, $\bm{E}$, including mutual polarization is,
\begin{equation}
  V(\bm{\mu}^{ind})=-\frac12\sum_i \bm{\mu}_i^{ind}\cdot \bm{E}_i^{overlap} + \sum_{i<j}\bm{\mu}^{ind}_i \bm{T}^{pol}_{ij,\mu\mu}\bm{\mu}^{ind}_j
  \label{eq:induced_dipoles}
\end{equation}
The field $\bm{E}_i^{overlap}$ is the damped electric field, and $\bm{T}^{pol}_{ij,\mu\mu}$ is the dipole-dipole interaction tensor which is derived from appropriate gradients of the polarization damping function $f_{ij}^{pol}/r_{ij}$. More specifically, the damping functions generated from Eq. \ref{eq:f_overlap} in models such as HIPPO\cite{rackers2021polarizable} have a polynomial that is one order too small to control the short-range singularity for mutual polarization, and we thus extend the polynomial by one order to formally eliminate polarization catastrophes. In addition, the fluctuating charge (FQ) contribution to the polarization energy is,
\begin{equation}
  V(\delta \bm{q})=\frac12\sum_i \eta_i \delta q_i^2 + \sum_i \delta q_i V_i + \sum_{i<j}\frac{\delta q_i \delta q_j}{r_{ij}} + \sum_{\alpha}\lambda_\alpha \sum_{i\in\alpha}\delta q_{i}
  \label{eq:fq}
\end{equation}
where $\eta_i$ is the atomic hardness of atom $i$, $\delta q$ are the optimally rearranged charges, and $\lambda$ are the Lagrange multipliers which enforce charge conservation. We determine the values of $\delta \bm{q}$ and $\bm{\mu}^{ind}$ by minimizing the total system energy by solving a system of linear equations.\cite{heindel2024} 

\textbf{Exchange Polarization}. We model exchange polarization, $V_{exch-pol}$, as
\begin{equation}
V_{exch,pol}=\sum_{i<j}\mathbf{M}_i^{exch,pol}\mathbf{T}_{ij}^{sr,-}\mathbf{M}_j^{exch,pol}
    \label{eq:exch_pol}
\end{equation}
\noindent
where $\mathbf{M}_i^{exch,pol}$ are the exchange-polarization multipoles which interact with one another via the attractive short-range multipole interaction tensor $\mathbf{T}_{ij}^{sr,-}$ defined in Eq. \ref{eq:mult_sr_minus}. We have written Eq. \ref{eq:exch_pol} in its multipolar form to emphasize its generality, but we only include the rank zero term for water and ions.

\textbf{Charge Transfer.} The CMM model for CT has both a direct and many-body contribution. The direct contributions allow for energetic stabilization associated with both forward and backward CT, and are described by an attractive short-range multipole expansion $\mathbf{T}_{ij}^{sr,-}$ as defined in Eq. \ref{eq:mult_sr_minus}.
\begin{subequations}
  \begin{equation}
V_{CT}^{i\rightarrow j}=\sum_{i<j}\mathbf{M}^{i\rightarrow j}_{CT} \mathbf{T}_{ij}^{sr,-}\mathbf{M}^{i\rightarrow j}_{CT}
\end{equation}
\begin{equation}
  V_{CT}^{j\rightarrow i}=\sum_{i<j}\mathbf{M}^{j\rightarrow i}_{CT} \mathbf{T}_{ij}^{sr,-}\mathbf{M}^{j\rightarrow i}_{CT}
\end{equation}
\begin{equation}
V_{CT}^{direct}=\sum_{i<j}V_{CT}^{i\rightarrow j}+V_{CT}^{j\rightarrow i}
\end{equation}
  \label{eq:ct_direct}
\end{subequations}

\noindent
For the CMM water model we choose to go up to rank two for the donor multipoles and rank zero for the acceptor multipoles. Since ions are spherically symmetric, they are only allowed rank zero donor multipoles. We allow charge to explicitly move between fragments by modifying the molecular charge constraints enforced while solving the set of linear equations for polarization. The amount of charge transferred between two molecules is parameterized to be proportional to the direct CT energy in Eq. \ref{eq:ct_direct}. By coupling CT into the polarization equations, we capture the so-called "repolarization" effect\cite{khaliullin2007} that gives rise to the indirect contributions to charge transfer,
\begin{equation}
  V^{indirect}_{CT}=V_{pol}(Q_{CT})-V_{pol}(0)
  \label{eq:ct_indirect}
\end{equation}
\noindent
The non-additive contribution to CT is defined as the polarization energy with CT, $V_{pol}(Q_{CT})$, minus the polarization energy without CT, $V_{pol}(0)$. 

\textbf{Reference Data.} The ion parameters are fit using ion-water configurations of size \ce{Z^{+/-}(H_2O)_n} with n=1-3. We fit parameters for \ce{Z^{+/-}}=\ce{F^-}, \ce{Cl^-}, \ce{Br^-}, \ce{I^-}, \ce{Li^+}, \ce{Na^+}, \ce{K^+}, \ce{Rb^+}, and \ce{Cs^+}. All energies and ALMO-EDA calculations used in fitting parameters of the CMM FF are computed at the $\omega$B97X-V level of DFT\cite{Mardirossian2014}, using the def2-QZVPPD basis set\cite{rappoport2010property}, and using the Q-Chem software package\cite{Epifanovsky2021}. 

For all ion-water pairs considered in this study, we ran a 10ps \textit{ab initio} molecular dynamics simulation at 500K with $\omega$B97X-V/def2-TZVPPD to generate probable ion-water configurations. We then sampled 2400 evenly spaced configurations from this trajectory to be used for parameterization. All larger ion-water clusters were generated by the following procedure. We used the Crest software package,\cite{pracht2020automated} which uses the semi-empirical GFN2-XTB\cite{bannwarth2019gfn2} method, to search for local minima on a potential energy surface. We carried out the Crest global minimum search with five different seed structures generated by taking water clusters, \ce{(H_2O)_n}, n=6-17, from a water cluster database\cite{rakshit2019atlas} and replacing one water randomly with one of the ions mentioned. We then took the structures of up to the ten lowest energy minima which had different hydrogen-bond networks and optimized them at the $\omega$B97X-V/def2-TZVPPD level of theory. This resulted in a total of 1044 unique ion-water clusters. These full clusters are used to characterize the ion-water potentials, but we also extracted all possible dimers and trimers from these clusters to be used in fitting of the ion force field parameters.

\textbf{Parameterization Procedure.} We fit each term against only the EDA contribution to that particular energy component. Optimization of parameters is done using simple gradient descent against the root mean-square deviation (RMSD) of predicted and EDA energies. For electrostatics and Pauli repulsion, we only use dimers in the fitting process since electrostatics is strictly pairwise-additive and Pauli repulsion is nearly so. For these terms, 200 random ion-water dimers from the datasets described above are used in fitting whereas for other many-body terms we use 200 random ion-water dimers, trimers, and tetramers from the datasets described above. After parameterizing all terms individually, the parameters relevant to Pauli repulsion are allowed to relax against interaction energies from the total dataset to improve error cancellation as described previously.\cite{heindel2024} 

\section*{Results}
\begin{figure}[H]
  \includegraphics[width=0.99\textwidth, trim={0 0 0 0},clip]{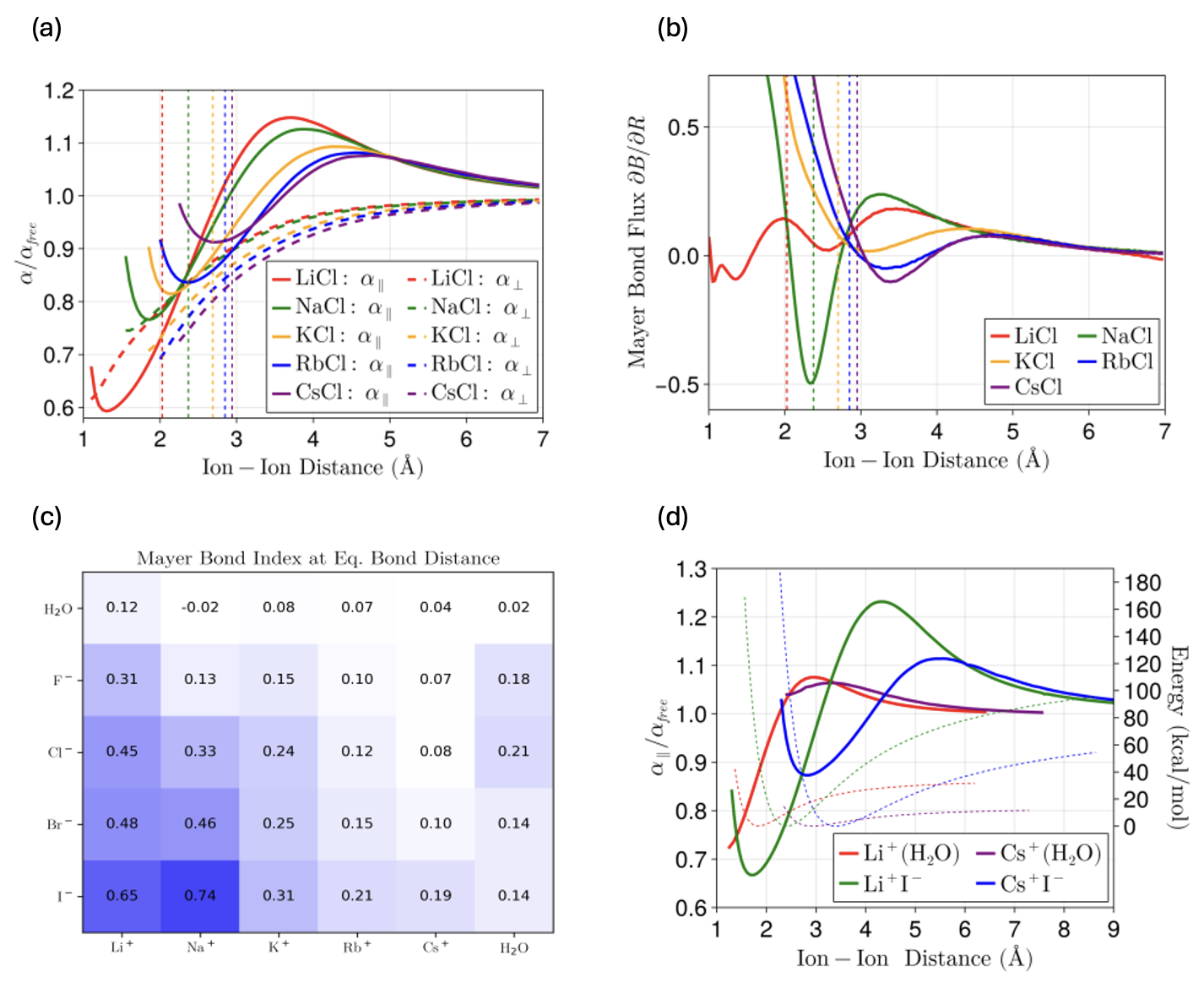}
  \caption{\textit{Evidence of partial covalency for ion-water and ion-ion interactions}. (a) Ratio of the pair polarizability and the sum of polarizabilities of the free ions on the ALMO surface. Using the ALMO surface allows for smooth dissociation to ionic fragments. Solid lines indicate the parallel component and dashed lines indicate the perpendicular component of the polarizability. (b) Derivative of the Mayer bond index with respect to distance for each ion pair. The polarizability ratio and bond flux mirror one another indicating their close connection, and confirming the partial covalency present at equilibrium. (c) Mayer bond index of all considered ion pairs at their equilibrium distance. Geometries and bond indices are computed with $\omega$B97X-V/def2-TZVPPD. (d) Polarizability ratio and energy of Li$^{+}-$water, Cs$^{+}-$water, Li$^{+}$-I$^{-}$, and Cs$^{+}$-I$^{-}$ as a function of distance. The potential energy is shown in dashed lines and is referenced to the axis on the right.}
  \label{fig:covalency}
\end{figure}

Figure \ref{fig:covalency}(a,b) show two  recent measures of full or partial covalency (or dative or ionic bonding) using the quantum mechanical polarizability metric\cite{hait2023bond} and the Mayer-Wiberg bond flux from wavefunction analysis\cite{Kim2024} for ion-ion pairing with chloride. There is a close connection between these quantities as they are both directly related to the total position spread of the electron density.\cite{hait2023bond,Kim2024} By either metric, a maximum along the bond dissociation coordinate defines a point which naturally separates the regime of shared electrons forming a partial bond from a regime corresponding to electron localization onto separate fragments. This partial covalency manifests by breaking the symmetry of the free atom polarizabilities and ultimately damping the atomic polarizabilities near the equilibrium distance of the ion interaction as shown in Figure \ref{fig:covalency}(a). It is clear from this analysis that the regime of (likely unequal) sharing of electrons occurs in the vicinity of the equilibrium distance of ion-ion interactions, and illustrates the difficulty for force fields to describe this inherent quantum mechanical interaction.

Figure \ref{fig:covalency}(c) shows the estimate of bond covalency from the Mayer bond index, and indicates that most covalency and hence largest polarizability anisotropy arise for ion pairs of small cations with large anions, while the opposite is true for ion pairs of large cations with small anions or with water. These observations are quite sensible considering that the electron binding energy of halide anions decreases with increasing size\cite{bouchafra2018predictive} while the electron affinity of alkali cations also decreases with size. 

Hence Figure \ref{fig:covalency}(d) analyzes the polarizability ratio parallel to the bond axis for \ce{Li^+} and \ce{Cs^+} paired with \ce{H2O} or \ce{I^-} as these correspond to the endpoints of the alkali/halide series and bracket the range of partial covalency character for the other ions. It is seen that larger cations interacting with water show minimal evidence of partial covalency as measured by the polarizability ratio. Specifically, the polarizability is strictly larger for \ce{Cs^+(H_2O)} than the sum of the separated species which indicates no electron localization. We will see that this is consistent with our ability to more accurately model \ce{Cs^+(H_2O)} polarization energies than those of \ce{Li^+(H_2O)}. In contrast,  \ce{Li^+(H_2O)} and \ce{Cs^+I^-} have similar amounts of partial covalency. However, the polarizability difference for \ce{Cs^+I^-} shows larger oscillations than \ce{Li^+(H_2O)} as shown in Supplementary Figure S1. This means that the energetic effect of partial covalency will be larger for ion pairs than for ion-water pairs. In this sense, we might interpret the polarizability ratio as quantifying the change in character of the lone pair involved in the dative bond while the polarizability difference better measures the energy associated with this change in electronic character.

The polarization modification which arises due to partial covalency in the ion interactions at short-range clearly must be addressed in an accurate many-body force field. The original AMOEBA model used constant polarizabilities which were lowered below their true values to improve their condensed phase properties.\cite{ren2011polarizable} More recently, the HIPPO model has incorporated polarizability damping proportional to the Slater density overlap\cite{chung2022classical} which is interpreted as arising from exchange-polarization. A note on the interpretation of terms is required here. In ALMO-EDA, the polarization energy naturally includes a contribution arising from the relief of Pauli repulsion.\cite{horn2016probing} This coupling arises because the polarization is a modification of the frozen energy term which contains the full Pauli repulsion. Therefore in CMM\cite{heindel2024}, we have defined exchange-polarization to be purely attractive as in Eq. \ref{eq:exch_pol} since we are actually modelling the decrease of Pauli repulsion associated with electronic polarization. In contrast, symmetry-adapted perturbation theory (SAPT) computes the full electrical response and then antisymmetrizes the system wavefunction resulting in a purely repulsive exchange-induction term.\cite{patkowski2020recent} This path-dependence in the energy is characteristic of energy decomposition schemes.\cite{andrada2020energy} We have argued in Figure \ref{fig:covalency} that the decrease of atomic polarizabilities under strong interactions arises from the onset of partial covalency. We prefer this more general interpretation, especially since, as discussed, exchange-polarization can also include attractive contributions depending on the choice of EDA scheme.

Therefore the atomic polarizabilities of interacting ions must be modified significantly compared to the free ions to account for electron rearrangements arising from partial covalency. The modified atomic polarizability must take into account the anisotropy between the parallel and perpendicular components of the polarizability, while near the equilibrium distance all directional components of the polarizability are decreased such that the molecular polarizability is $\sim$10\% to $\sim$30\% smaller than the sum of the gas-phase polarizabilities of each ion depending on the ion pair.

We thus formulate an environment-dependent change in the polarizability parameter of an ion, $\alpha(\mathbf{E})$
\begin{equation}
  \alpha(\mathbf{E})=\alpha_0-\alpha_0\lambda_{max}\left(1-e^{-b_E \mathbf{E}\cdot \mathbf{E}}\right)
  \label{eq:pol_damping}
\end{equation}
\noindent
by making it a function of the electric field, $\mathbf{E}$. This choice avoids the use of any pair-specific or distance-dependent parameters; instead, the strength of the interaction is encoded in the strength of the field. It is important to note that this modification of the polarizability only becomes relevant when the local electric field approaches $\approx 150$ MV/cm which makes it primarily relevant to ion pairing. The parameter $\lambda_{max}$ is a number between 0 and 1 which enforces a lower-bound on the polarizability, a behavior consistent with calculations from electronic structure. This modification to polarization completes the definition of the CMM model for ions and is a central result of this work.

\begin{figure}[h]
  
  \begin{subfigure}{0.49\textwidth}
    \includegraphics*[width=\textwidth]{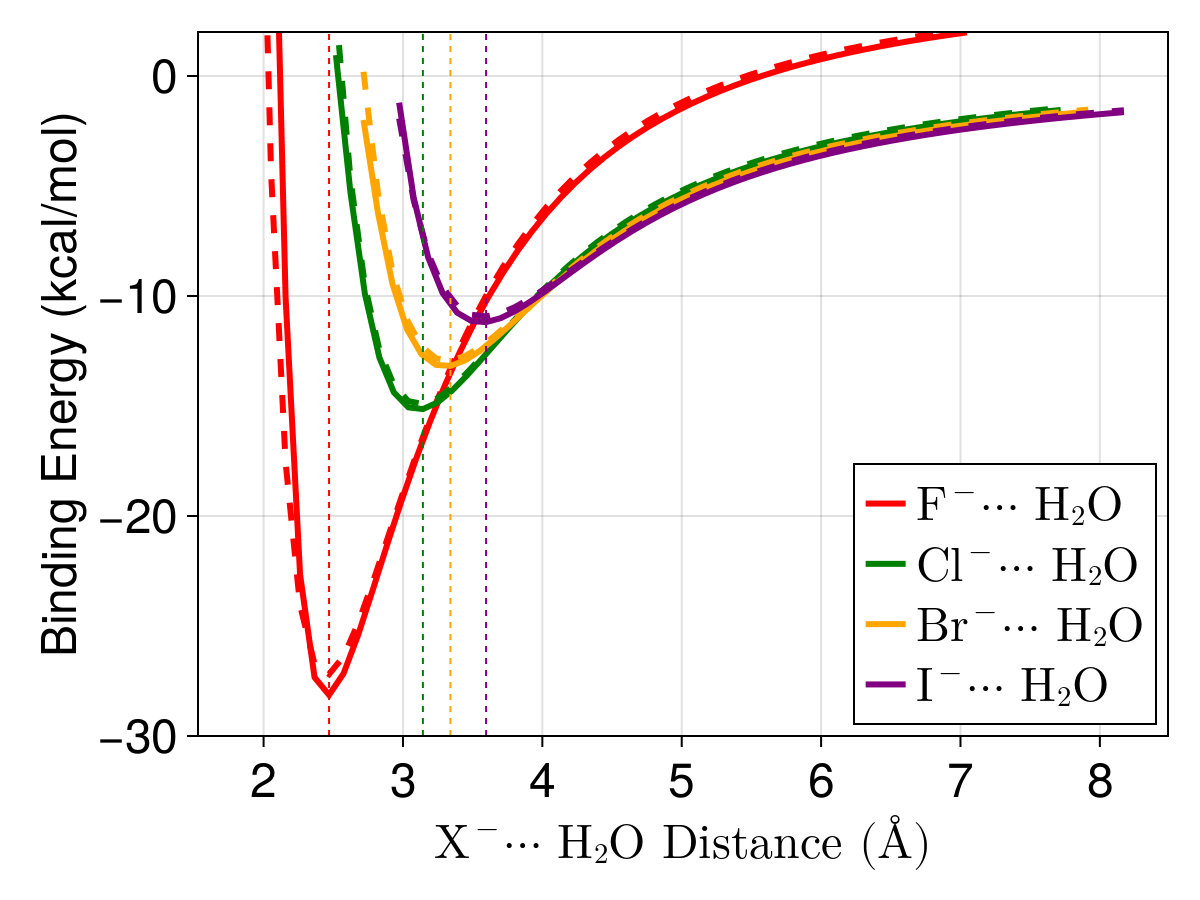}
  \end{subfigure}
  \begin{subfigure}{0.49\textwidth}
    \includegraphics*[width=\textwidth]{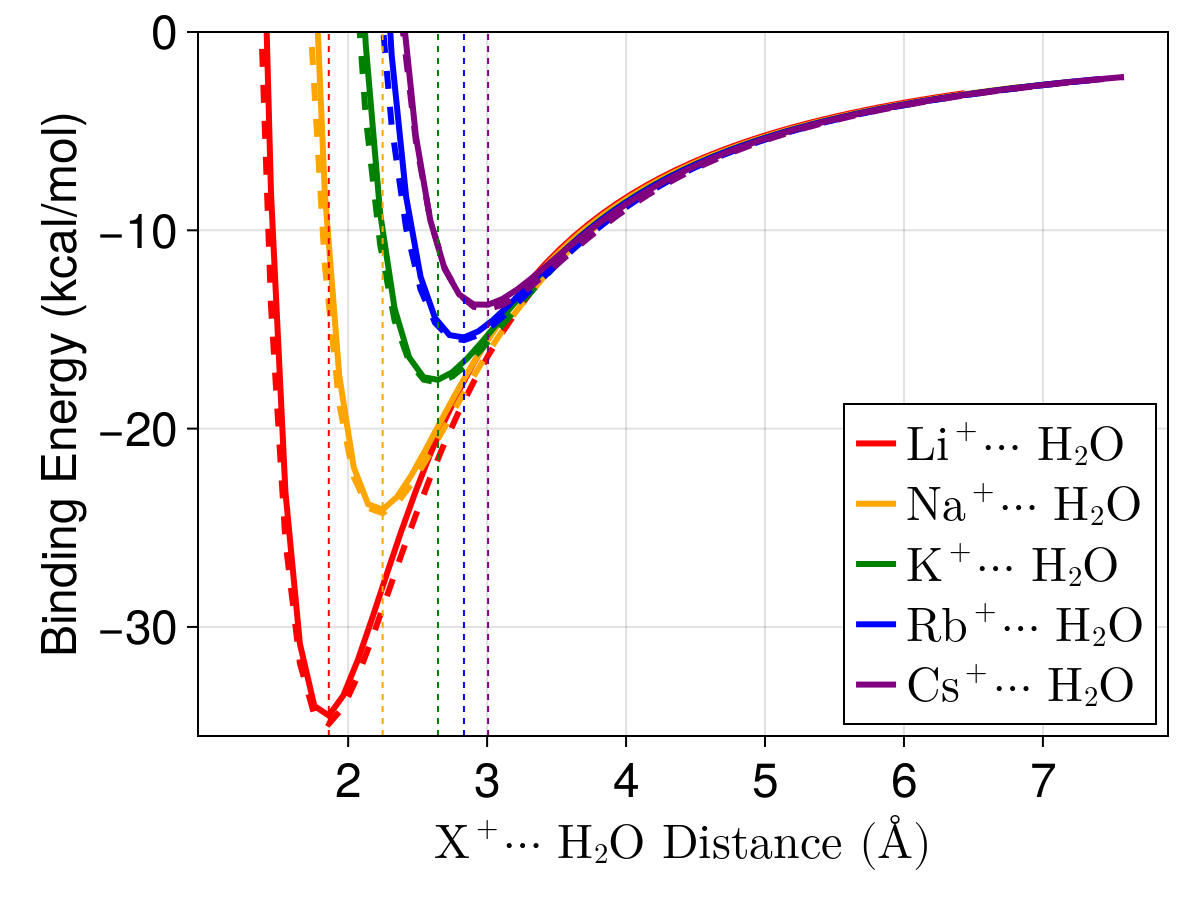}
  \end{subfigure}
  \caption{\textit{Total energy scans for water-ion dimers as a function of distance.} (A) anions (\ce{X^-}=\ce{F^-}, \ce{Cl^-}, \ce{Br^-}, and \ce{I^-}) and (B) cations (\ce{X^+}=\ce{Li^+}, \ce{Na^+}, \ce{K^+}, \ce{Rb^+}, and \ce{Cs^+}). The water geometry is held fixed at the equilibrium structure of each dimer. The solid curve corresponds to CMM and the dashed curve corresponds to $\omega$B97X-V/def2-QZVPPD.  Vertical dashed lines show the positions of each minimum with $\omega$B97X-V/def2-QZVPPD.
  Binding energies and harmonic frequencies are reported in Table \ref{tab:ion_freqs}.
  }\label{fig:ion_scans}
\end{figure}

Figure \ref{fig:ion_scans} provides the first test of the CMM model for ion-water interactions by evaluating the total CMM two-body energy against the DFT reference energy. It is evident that CMM captures the long-range attractive part of the potential very accurately in all cases, but more importantly reproduces quite well the interactions in the overlapping region at short-range. The largest error is found for \ce{F^-(H_2O)} (1.0 kcal/mol), while the rest of the ions have binding energies accurate to within a few tenths of a kcal/mol. This is unsurprising since \ce{F^-} and \ce{Li^+} are the most challenging ions considered in this work due to their extremely short-range and stronger molecular interactions (Figure \ref{fig:ion_scans}). 

\begin{figure}[H]
  \includegraphics[width=0.9\textwidth]{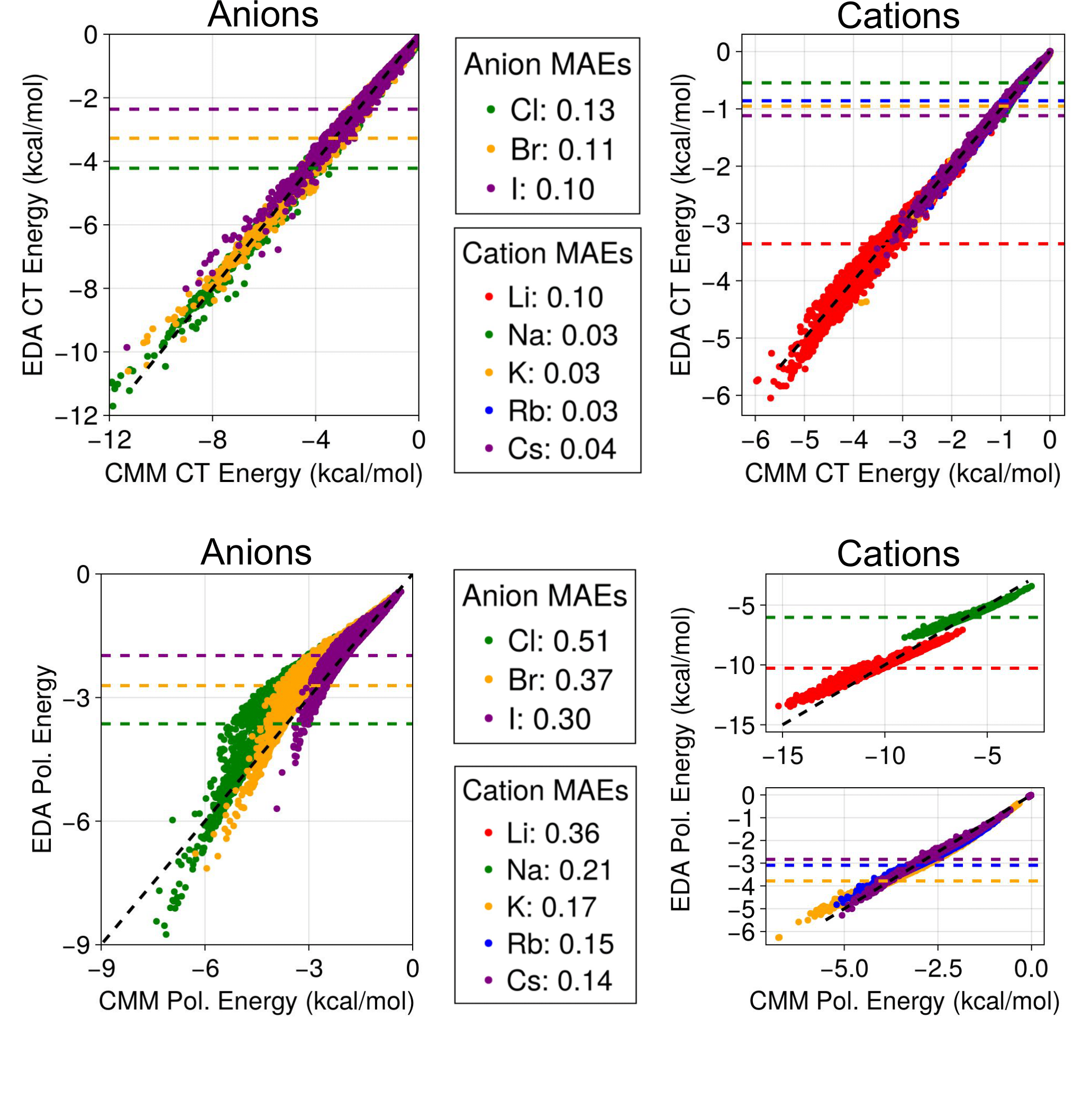}
  \caption{\textit{Two-body polarization and charge transfer energies between CMM and EDA for anion-water and cation-water interactions.} All dimers are drawn from $\omega$B97X-V/def2-TZVPPD AIMD simulations with EDA energies computed with $\omega$B97X-V/def2-QZVPPD. The MAES of all ion-water interactions are reported in the legend in kcal/mol.
}
  \label{fig:Pol_CT_ions}
\end{figure}

Supplementary Figures S2 and S3 provide the errors in each individual EDA component of the scan for water interacting with halide anions and alkali metal cations, respectively. Overall the errors against each EDA term are less than 1 kcal/mol with all terms becoming exact at around 7-8 \AA. With the help of error cancellation among the energy components at very short-range, an accurate ion-water profile in the strongly overlapping regime is achieved. Figure \ref{fig:Pol_CT_ions} shows that while the two-body charge transfer is mostly unbiased for all cations and anions considered, with MAEs of 0.10 to 0.13 kcal/mol for each anion and between 0.03 and 0.10 kcal/mol for the cations, there are some systematic errors that become more pronounced as the ions get smaller for polarization. This might indicate a need for quadrupolar (or even higher-order) polarization terms\cite{buckingham1979polarizability,wilson1996quadrupole} for ion-water interactions, but it appears that the atomic polarization modification introduced in Eq. \ref{eq:pol_damping} is able to adequately reproduce the EDA polarization, since the MAEs are between 0.14 and 0.51 kcal/mol.

\begin{figure}[H]
    \includegraphics*[width=0.9\textwidth]{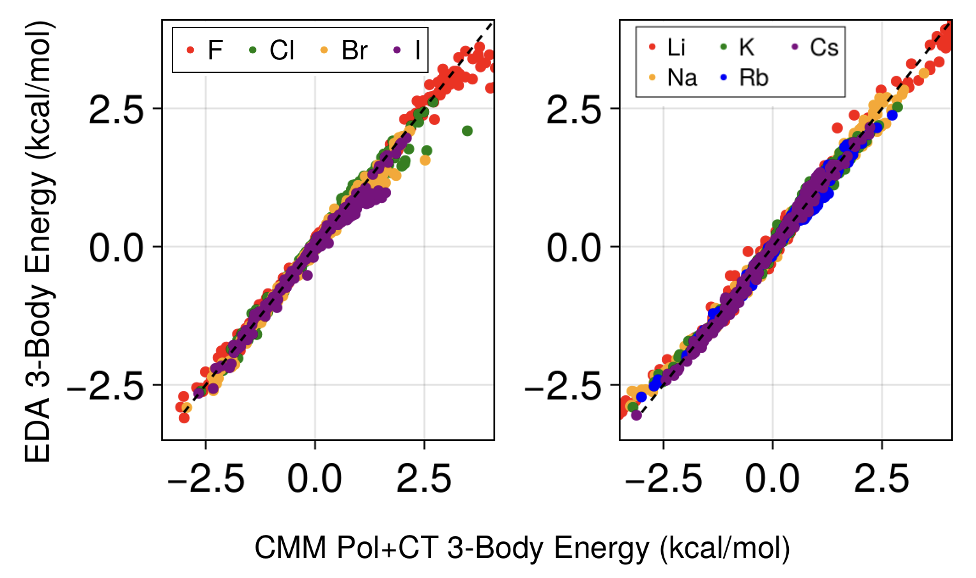}
  \caption{\textit{Correlation of summed three-body polarization and charge transfer energies between CMM and EDA for anions and cations.} All trimers are drawn from ion-water clusters optimized with $\omega$B97X-V/def2-TZVPPD and energies are computed with the def2-QZVPPD basis set. The 3-body MAES of the \ce{F^-}, \ce{Cl^-}, \ce{Br^-}, and \ce{I^-} water-ion trimers for are 0.092, 0.059, 0.054, and 0.044 kcal/mol. The 3-body MAES of the \ce{Li^+}, \ce{Na^+}, \ce{K^+}, \ce{Rb^+}, and \ce{Cs^+} water-ion trimers are 0.141, 0.071, 0.070, 0.075, and 0.065 kcal/mol.
}
  \label{fig:3BPol_CT_ions}
\end{figure}

Another important feature of the ion-water potential in the condensed phase is that many-body effects shorten the oxygen-ion distance relative to the dimer equilibrium distance in the gas phase. In Figure \ref{fig:3BPol_CT_ions} we find that the summed three-body polarization and charge transfer energies computed from CMM for both water-anion and water-cation trimers capture the corresponding summed three-body EDA terms very accurately. The correlations between three-body polarization and charge transfer are shown separately in Supplementary Figure S4, in which we observe (for anions in particular), there are compensating errors in the three-body polarization and charge transfer. The fact the errors in polarization and charge transfer are inversely correlated arises naturally from the coupling of polarization and charge transfer described in Eq. \ref{eq:ct_indirect}, and aid in error cancellation. We note that in past work, large polarizable anions like \ce{I^-} have presented a problem for polarizable force fields, but CMM does not have any problem with large polarizabilities due to the choice of an appropriate damping function for polarization as described in our previous CMM for water.\cite{heindel2024} 

\begin{table}[H]
  \begin{center}
  \begin{tabular}{llccccccc}
      \multicolumn{9}{c}{Ion-Water Dimer Vibrational Frequencies} \\\hline
      \ce{X^{+/-}(H2O)} & Method & NM1 & NM2 & NM3 & NM4 & NM5 & NM6 & $D_e$ \\\hline
      \ce{F^-(H2O)} & CMM   & 426 & 672 & 1368 & 1826 & 2343 & 3930 & -28.2 \\
           & $\omega$B97X-V & 384 & 569 & 1144 & 1702 & 2233 & 3916 & -27.2 \\\hline
      \ce{Cl^-(H2O)} & CMM  & 204 & 362 & 796 & 1714 & 3417 & 3931 & -15.2 \\
           & $\omega$B97X-V & 193 & 341 & 726 & 1678 & 3417 & 3919 & -14.9 \\\hline
      \ce{Br^-(H2O)} & CMM  & 159 & 310 & 721 & 1705 & 3528 & 3923 & -13.3 \\
           & $\omega$B97X-V & 151 & 288 & 657 & 1673 & 3521 & 3916 & -12.9 \\\hline
      \ce{I^-(H2O)} & CMM   & 133 & 221 & 629 & 1689 & 3590 & 3905 & -11.3 \\
           & $\omega$B97X-V & 118 & 220 & 579 & 1668 & 3623 & 3911 & -10.9 \\\hline
      \ce{Li^+(H2O)} & CMM  & 346 & 549 & 609 & 1688 & 3626 & 3769 & -34.7 \\
           & $\omega$B97X-V & 392 & 524 & 554 & 1681 & 3815 & 3882 & -34.9 \\\hline
      \ce{Na^+(H2O)} & CMM  & 260 & 334 & 426 & 1685 & 3713 & 3833 & -24.1 \\
           & $\omega$B97X-V & 307 & 367 & 437 & 1677 & 3830 & 3902 & -24.3 \\\hline
      \ce{K^+(H2O)} & CMM   & 230 & 294 & 369 & 1683 & 3752 & 3863 & -17.6 \\
           & $\omega$B97X-V & 213 & 359 & 369 & 1673 & 3833 & 3910 & -17.7 \\\hline
      \ce{Rb^+(H2O)} & CMM  & 191 & 295 & 348 & 1681 & 3764 & 3873 & -15.5 \\
           & $\omega$B97X-V & 178 & 347 & 351 & 1671 & 3836 & 3914 & -15.6 \\\hline
      \ce{Cs^+(H2O)} & CMM  & 165 & 290 & 333 & 1680 & 3774 & 3881 & -13.9 \\
           & $\omega$B97X-V & 157 & 327 & 339 & 1668 & 3836 & 3916 & -14.0 \\\hline
  \end{tabular}
  \end{center}
  \vspace{-3mm}
  \caption{\textit{Comparison of the CMM dimer vibrational frequencies and binding energies against $\omega$B97X-V/def2-QZVPPD for all alkali and halogen ion-water dimers.} Frequency and energy units are $\mathrm{cm^{-1}}$ and kcal/mol, respectively.}
  \label{tab:ion_freqs}
\end{table}

\vspace{-2mm}
Since ion-water interactions are strong, they can result in large shifts of the underlying water vibrational frequencies. Table \ref{tab:ion_freqs} provides the frequencies and binding energies of CMM compared to $\omega$B97X-V/def2-QZVPPD in which the CMM is generally quite accurate for the high frequency free \ce{O-H} (NM6) for all ions, and the stretching mode (NM5) of the larger cations and anions. The water bending mode for cation-water dimers, NM4, is very accurate while that for anions is reasonably accurate albeit systematically blue-shifted. This is because the anion attracts the water hydrogen atom causing the \ce{HOH} angle to close more than it should. This problem could be corrected by incorporating a field-dependent bending potential\cite{lacour2023predicting} but this is left for future work. The low-frequency vibrations are reproduced very accurately which is a good indication that CMM is able to model the solvent rearrangements relevant to ion diffusion.

\begin{figure}[H]
  \includegraphics*[width=0.9\textwidth]{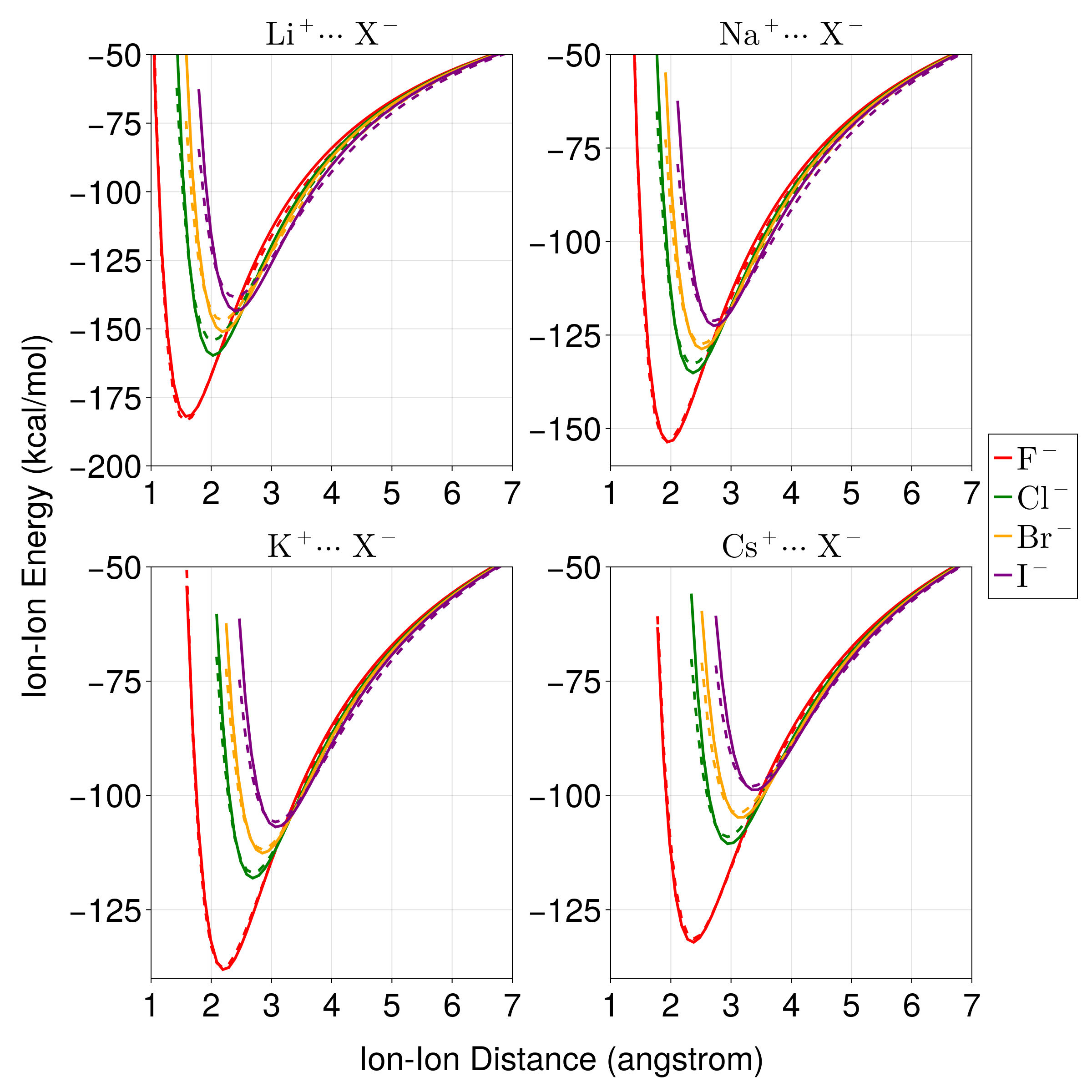}
  \caption{\textit{Scans of total energies for alkali-halide ions pairs.} Total energies associated with ion pairing for all pairs of \ce{Li^+}, \ce{Na^+}, \ce{K^+}, and \ce{Cs^+} with \ce{F^-}, \ce{Cl^-}, \ce{Br^-}, and \ce{I^-}.
}
  \label{fig:ion_pairs}
\end{figure}

\vspace{-2mm}
As ion pairing interactions are much stronger than ion-water interactions the issue of partial covalency is even more pronounced. Despite these difficulties, we are able to achieve rather accurate ion-pairing profiles as shown in Figure \ref{fig:ion_pairs}. Similar scans which show the error in individual EDA components for all considered ions are shown in Supplementary Figures S5-S9. As with ion-water interactions, the larger the ions involved, the smaller the errors in the individual EDA terms. The only exception is for polarization energies where larger anions tend to result in slightly larger errors due to increased covalency of the interaction.

\begin{figure}[H]
  \includegraphics*[width=0.9\textwidth]{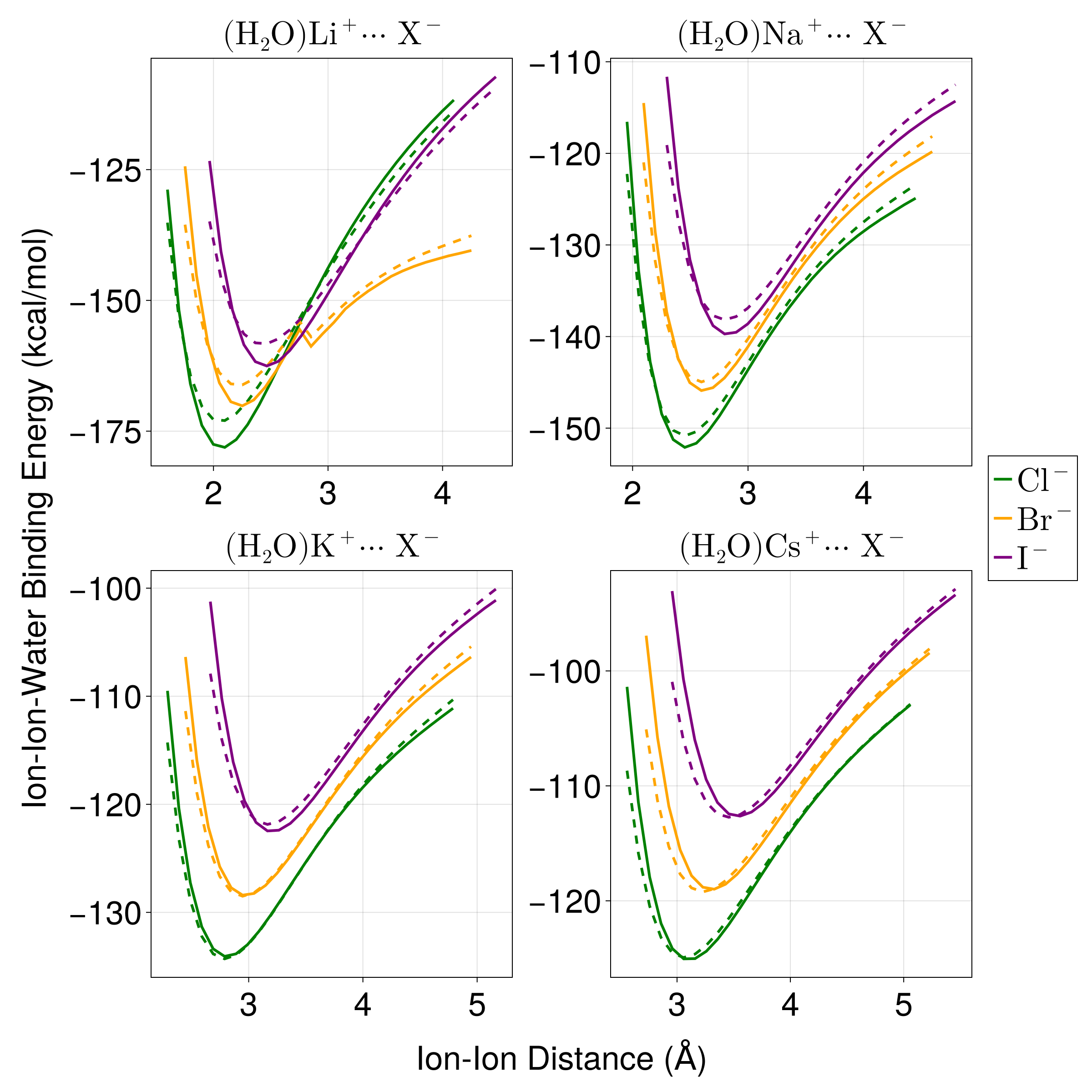}
  \caption{\textit{Relaxed scans of total energies for alkali-halide solvent-separated ion pairs.} Total energies for \ce{H_2O(Z^+\cdots X^-)} where \ce{Z^+}=\ce{Li^+}, \ce{Na^+}, \ce{K^+}, \ce{Cs^+} and \ce{X^-}=\ce{Cl^-}, \ce{Br^-}, \ce{I^-}. \ce{F^-} is omitted since it results in dissociation of the water molecule which is outside the scope of this non-reactive model.
}
  \label{fig:ssip_scans}
\end{figure}

We now turn our attention to many-body interactions between water and multiple ions. We begin by testing the performance of the model for a relaxed scan of the solvent-separated ion pairs (SSIPs) of alkali-halide ion pairs with water. In Figure \ref{fig:ssip_scans}, we see that the agreement between CMM and $\omega$B97X-V ranges from fair to very good depending on the distance and ion pair considered. For larger cations, the errors are generally quite small especially considering the large interaction energies. This is consistent with our discussion of the diminished amount of covalency present when larger cations are involved in ion pairing.

\begin{figure}[H]
  \includegraphics*[width=\textwidth]{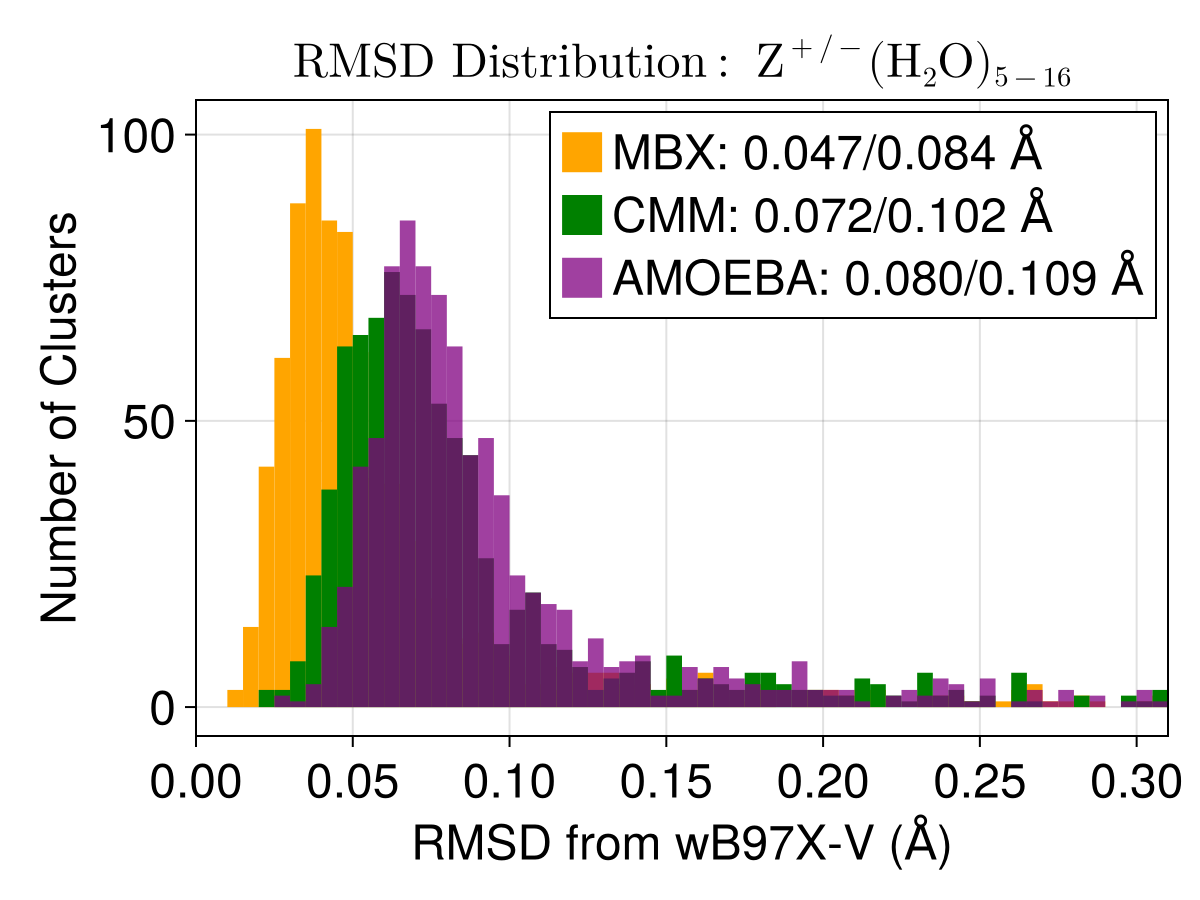}
  \caption{\textit{Histogram of all ion-water RMSDs against $\omega$B97X-V/def2-TZVPPD.} In total, 1044 clusters of \ce{Z^{+/-}(H_2O)_{5-16}} where \ce{Z^+} are alkali cations and \ce{Z^-} are halide anions were optimized with each potential. The RMSD of each force field optimized structure is computed against $\omega$B97X-V resulting in the above histograms. The legend shows the median RMSD followed by the average RMSD. See text for discussion.}
  \label{fig:ion_water_rmsds}
\end{figure}

Another important measure of the quality of a potential is its ability to reproduce the structure of local solvation environments. In previous work, we have shown that CMM reproduces benchmark water cluster geometries with very high accuracy, even surpassing the very accurate fitted potential MB-Pol on this particular measure.\cite{heindel2024} We carry out a similar analysis in Figure \ref{fig:ion_water_rmsds} on an extensive dataset of ion water geometries, \ce{Z^{+/-}(H_2O)_{5-16}}, all optimized with $\omega$B97X-V/def2-TZVPPD, and make a comparison to MBX\cite{riera2017toward,caruso2021data,caruso2022accurate} and AMOEBA\cite{ponder2010current} (using the amoebabio18 parameter set)\cite{zhang2018amoeba}. 

MBX, CMM, and AMOEBA all reproduce the structures of ion-water clusters with satisfactory accuracy, but there are clear differences in performance among the models. When optimizing to the correct QM isomer (RMSD $\lessapprox$ 0.125 \AA), MBX performs slightly better than CMM since it explicitly uses a high polynomial fit to CCSD(T)-F12b\cite{werner2011explicitly} dimer and trimer energies. Hence although CMM outperforms MB-Pol (the underlying water model of MBX) in reproducing pure water cluster geometries\cite{heindel2024}, and many-body CMM errors are small, we attribute the gap between MBX and CMM structure performance largely to errors in the 2-body contribution to ion-water interactions. Correspondingly, AMOEBA performs slightly worse than CMM since, although an advanced force field, lacks CT and other advanced functional forms used in the CMM. Note that all clusters with RMSDs larger than $\approx$ 0.125 \AA~ have collapsed to a slightly different isomer than the reference due to model errors in the forces in the initial geometry, an effect that is an issue even for MBX. Rather than separating these outliers from the clusters which remained stable, we simply report the median and average RMSDs over the entire data set. The median and mean RMSDs for each model are broken down by ion in Supplementary Table S1 which shows fairly consistent performance and errors for all ions. 

\section*{Discussion and Conclusions}
Accurate force fields for strong ion-water and ion-ion interactions have long been a challenge for molecular simulation. Non-polarizable force fields are intrinsically unable to model large many-body energies, while polarizable force fields such as AMOEBA have recognized the need to use atomic polarizabilities which are different than their gas-phase values for ion interactions\cite{ponder2010current}. While recent polarizable models such as HIPPO\cite{chung2022classical} have introduced atomic ion polarizabilities that decrease with interaction distance, we think it is more accurate to refer to this effect as arising from partial covalency rather than exchange-polarization specifically.

Instead it is an underappreciated fact that ion-ion and some ion-water interactions include a non-negligible covalent contribution at equilibrium. As recently pointed out by Hait and M. Head-Gordon, the polarizability is a good reporter on the covalency of an interaction.\cite{hait2023bond} In short, a polarizability maximum along the atomic distance can be reasonably defined as the point of bond-breaking since this is the point separating charge localization on each atom from charge localization between the atoms. Kim and T. Head-Gordon have used the Mayer (and Wiberg) bond indices to also elicit evidence of partial covalency for ion-ion and ion-water interactions. Both metrics show significant evidence of dative or ionic bonding at equilibrium for ion interactions, illustrating why force fields have struggled with such systems previously, but also have helped us to trace back through EDA how to account for the partial covalency. 

We therefore have introduced a new model for ion atomic polarizabilities that modifies their magnitude based on their local environment as assessed by the electric field. The modification of ion polarizabilities by their environment eliminates the need for ion specific parameters, and is physically motivated by its importance for condensed phases involving crystals\cite{mahan1980polarizability}, molten salts,\cite{jacucci1976effects} interactions with water\cite{pyper1992polarizabilities,li2017accurate}, and now for dative or ionic bonding.  Given the underappreciated fact that ion-ion and some ion-water interactions include a non-negligible covalent contribution at equilibrium, we believe that the CMM shows excellent results for ion-water interactions, and good to excellent results for ion-ion pairing depending on ion size and degree of partial covalency. In future work we will extend the validation suite of CMM for ions by examining ion hydration free energies and transport properties in the condensed phase.

\begin{acknowledgement}
We acknowledge support from the U.S. National Science Foundation through Grant No. CHE-2313791. Computational resources were provided by the National Energy Research Scientific Computing Center (NERSC), a U.S. Department of Energy Office of Science User Facility operated under Contract DE-AC02-05CH11231. 
\end{acknowledgement}

\begin{suppinfo}
Complete 2-body and 3-body EDA analysis for all ions considered.
\end{suppinfo}

\bibliography{references}

\providecommand{\latin}[1]{#1}
\makeatletter
\providecommand{\doi}
  {\begingroup\let\do\@makeother\dospecials
  \catcode`\{=1 \catcode`\}=2 \doi@aux}
\providecommand{\doi@aux}[1]{\endgroup\texttt{#1}}
\makeatother
\providecommand*\mcitethebibliography{\thebibliography}
\csname @ifundefined\endcsname{endmcitethebibliography}  {\let\endmcitethebibliography\endthebibliography}{}
\begin{mcitethebibliography}{48}
\providecommand*\natexlab[1]{#1}
\providecommand*\mciteSetBstSublistMode[1]{}
\providecommand*\mciteSetBstMaxWidthForm[2]{}
\providecommand*\mciteBstWouldAddEndPuncttrue
  {\def\EndOfBibitem{\unskip.}}
\providecommand*\mciteBstWouldAddEndPunctfalse
  {\let\EndOfBibitem\relax}
\providecommand*\mciteSetBstMidEndSepPunct[3]{}
\providecommand*\mciteSetBstSublistLabelBeginEnd[3]{}
\providecommand*\EndOfBibitem{}
\mciteSetBstSublistMode{f}
\mciteSetBstMaxWidthForm{subitem}{(\alph{mcitesubitemcount})}
\mciteSetBstSublistLabelBeginEnd
  {\mcitemaxwidthsubitemform\space}
  {\relax}
  {\relax}

\bibitem[Jungwirth and Tobias(2006)Jungwirth, and Tobias]{Jungwirth2006}
Jungwirth,~P.; Tobias,~D.~J. Specific Ion Effects at the Air/Water Interface. \emph{Chemical Reviews} \textbf{2006}, \emph{106}, 1259--1281\relax
\mciteBstWouldAddEndPuncttrue
\mciteSetBstMidEndSepPunct{\mcitedefaultmidpunct}
{\mcitedefaultendpunct}{\mcitedefaultseppunct}\relax
\EndOfBibitem
\bibitem[Mamatkulov and Schwierz(2018)Mamatkulov, and Schwierz]{Mamatkulov2018}
Mamatkulov,~S.; Schwierz,~N. Force fields for monovalent and divalent metal cations in TIP3P water based on thermodynamic and kinetic properties. \emph{The Journal of Chemical Physics} \textbf{2018}, \emph{148}, 074504\relax
\mciteBstWouldAddEndPuncttrue
\mciteSetBstMidEndSepPunct{\mcitedefaultmidpunct}
{\mcitedefaultendpunct}{\mcitedefaultseppunct}\relax
\EndOfBibitem
\bibitem[Xantheas(2000)]{xantheas2000cooperativity}
Xantheas,~S.~S. Cooperativity and hydrogen bonding network in water clusters. \emph{Chemical Physics} \textbf{2000}, \emph{258}, 225--231\relax
\mciteBstWouldAddEndPuncttrue
\mciteSetBstMidEndSepPunct{\mcitedefaultmidpunct}
{\mcitedefaultendpunct}{\mcitedefaultseppunct}\relax
\EndOfBibitem
\bibitem[Heindel and Xantheas(2021)Heindel, and Xantheas]{heindel2021many}
Heindel,~J.~P.; Xantheas,~S.~S. The many-body expansion for aqueous systems revisited: II. Alkali metal and halide ion--water interactions. \emph{Journal of chemical theory and computation} \textbf{2021}, \emph{17}, 2200--2216\relax
\mciteBstWouldAddEndPuncttrue
\mciteSetBstMidEndSepPunct{\mcitedefaultmidpunct}
{\mcitedefaultendpunct}{\mcitedefaultseppunct}\relax
\EndOfBibitem
\bibitem[Herman \latin{et~al.}(2021)Herman, Heindel, and Xantheas]{herman2021many}
Herman,~K.~M.; Heindel,~J.~P.; Xantheas,~S.~S. The many-body expansion for aqueous systems revisited: III. Hofmeister ion--water interactions. \emph{Physical Chemistry Chemical Physics} \textbf{2021}, \emph{23}, 11196--11210\relax
\mciteBstWouldAddEndPuncttrue
\mciteSetBstMidEndSepPunct{\mcitedefaultmidpunct}
{\mcitedefaultendpunct}{\mcitedefaultseppunct}\relax
\EndOfBibitem
\bibitem[Joung and Cheatham(2008)Joung, and Cheatham]{Joung2008}
Joung,~I.~S.; Cheatham,~I.,~Thomas~E. Determination of Alkali and Halide Monovalent Ion Parameters for Use in Explicitly Solvated Biomolecular Simulations. \emph{The Journal of Physical Chemistry B} \textbf{2008}, \emph{112}, 9020--9041\relax
\mciteBstWouldAddEndPuncttrue
\mciteSetBstMidEndSepPunct{\mcitedefaultmidpunct}
{\mcitedefaultendpunct}{\mcitedefaultseppunct}\relax
\EndOfBibitem
\bibitem[Leontyev and Stuchebrukhov(2011)Leontyev, and Stuchebrukhov]{leontyev2011accounting}
Leontyev,~I.; Stuchebrukhov,~A. Accounting for electronic polarization in non-polarizable force fields. \emph{Physical Chemistry Chemical Physics} \textbf{2011}, \emph{13}, 2613--2626\relax
\mciteBstWouldAddEndPuncttrue
\mciteSetBstMidEndSepPunct{\mcitedefaultmidpunct}
{\mcitedefaultendpunct}{\mcitedefaultseppunct}\relax
\EndOfBibitem
\bibitem[Bedrov \latin{et~al.}(2019)Bedrov, Piquemal, Borodin, MacKerell~Jr, Roux, and Schroeder]{bedrov2019molecular}
Bedrov,~D.; Piquemal,~J.-P.; Borodin,~O.; MacKerell~Jr,~A.~D.; Roux,~B.; Schroeder,~C. Molecular dynamics simulations of ionic liquids and electrolytes using polarizable force fields. \emph{Chemical reviews} \textbf{2019}, \emph{119}, 7940--7995\relax
\mciteBstWouldAddEndPuncttrue
\mciteSetBstMidEndSepPunct{\mcitedefaultmidpunct}
{\mcitedefaultendpunct}{\mcitedefaultseppunct}\relax
\EndOfBibitem
\bibitem[Mao \latin{et~al.}(2017)Mao, Horn, and Head-Gordon]{mao2017energy}
Mao,~Y.; Horn,~P.~R.; Head-Gordon,~M. Energy decomposition analysis in an adiabatic picture. \emph{Physical Chemistry Chemical Physics} \textbf{2017}, \emph{19}, 5944--5958\relax
\mciteBstWouldAddEndPuncttrue
\mciteSetBstMidEndSepPunct{\mcitedefaultmidpunct}
{\mcitedefaultendpunct}{\mcitedefaultseppunct}\relax
\EndOfBibitem
\bibitem[Demerdash \latin{et~al.}(2017)Demerdash, Mao, Liu, Head-Gordon, and Head-Gordon]{demerdash2017assessing}
Demerdash,~O.; Mao,~Y.; Liu,~T.; Head-Gordon,~M.; Head-Gordon,~T. Assessing many-body contributions to intermolecular interactions of the AMOEBA force field using energy decomposition analysis of electronic structure calculations. \emph{The Journal of chemical physics} \textbf{2017}, \emph{147}, 161721\relax
\mciteBstWouldAddEndPuncttrue
\mciteSetBstMidEndSepPunct{\mcitedefaultmidpunct}
{\mcitedefaultendpunct}{\mcitedefaultseppunct}\relax
\EndOfBibitem
\bibitem[Heindel \latin{et~al.}(2024)Heindel, Sami, and Head-Gordon]{heindel2024}
Heindel,~J.~P.; Sami,~S.; Head-Gordon,~T. Completely Multipolar Model as a General Framework for Many-Body Interactions as Illustrated for Water. 2024; \url{https://arxiv.org/abs/2406.15944}\relax
\mciteBstWouldAddEndPuncttrue
\mciteSetBstMidEndSepPunct{\mcitedefaultmidpunct}
{\mcitedefaultendpunct}{\mcitedefaultseppunct}\relax
\EndOfBibitem
\bibitem[Khaliullin \latin{et~al.}(2007)Khaliullin, Cobar, Lochan, Bell, and Head-Gordon]{khaliullin2007}
Khaliullin,~R.~Z.; Cobar,~E.~A.; Lochan,~R.~C.; Bell,~A.~T.; Head-Gordon,~M. Unravelling the origin of intermolecular interactions using absolutely localized molecular orbitals. \emph{The Journal of Physical Chemistry A} \textbf{2007}, \emph{111}, 8753--8765\relax
\mciteBstWouldAddEndPuncttrue
\mciteSetBstMidEndSepPunct{\mcitedefaultmidpunct}
{\mcitedefaultendpunct}{\mcitedefaultseppunct}\relax
\EndOfBibitem
\bibitem[Horn \latin{et~al.}(2016)Horn, Mao, and Head-Gordon]{horn2016probing}
Horn,~P.~R.; Mao,~Y.; Head-Gordon,~M. Probing non-covalent interactions with a second generation energy decomposition analysis using absolutely localized molecular orbitals. \emph{Physical Chemistry Chemical Physics} \textbf{2016}, \emph{18}, 23067--23079\relax
\mciteBstWouldAddEndPuncttrue
\mciteSetBstMidEndSepPunct{\mcitedefaultmidpunct}
{\mcitedefaultendpunct}{\mcitedefaultseppunct}\relax
\EndOfBibitem
\bibitem[Mao \latin{et~al.}(2021)Mao, Loipersberger, Horn, Das, Demerdash, Levine, Veccham, {Head-Gordon}, and {Head-Gordon}]{Mao:2021:EDA-review}
Mao,~Y.; Loipersberger,~M.; Horn,~P.~R.; Das,~A.; Demerdash,~O.; Levine,~D.~S.; Veccham,~S.~P.; {Head-Gordon},~T.; {Head-Gordon},~M. From intermolecular interaction energies and observable shifts to component contributions and back again: {A} tale of variational energy decomposition analysis. \textbf{2021}, \emph{72}, 641--666\relax
\mciteBstWouldAddEndPuncttrue
\mciteSetBstMidEndSepPunct{\mcitedefaultmidpunct}
{\mcitedefaultendpunct}{\mcitedefaultseppunct}\relax
\EndOfBibitem
\bibitem[Hait and Head-Gordon(2023)Hait, and Head-Gordon]{hait2023bond}
Hait,~D.; Head-Gordon,~M. When is a bond broken? The polarizability perspective. \emph{Angewandte Chemie International Edition} \textbf{2023}, \emph{62}, e202312078\relax
\mciteBstWouldAddEndPuncttrue
\mciteSetBstMidEndSepPunct{\mcitedefaultmidpunct}
{\mcitedefaultendpunct}{\mcitedefaultseppunct}\relax
\EndOfBibitem
\bibitem[Kim and Head-Gordon(2024)Kim, and Head-Gordon]{Kim2024}
Kim,~L.; Head-Gordon,~T. Near Equivalence of Polarizability and Bond Order Flux Metrics for Describing Covalent Bond Rearrangements. 2024; \url{https://arxiv.org/abs/2408.14643}\relax
\mciteBstWouldAddEndPuncttrue
\mciteSetBstMidEndSepPunct{\mcitedefaultmidpunct}
{\mcitedefaultendpunct}{\mcitedefaultseppunct}\relax
\EndOfBibitem
\bibitem[Boyer \latin{et~al.}(2019)Boyer, Marsalek, Heindel, Markland, McCoy, and Xantheas]{boyer2019beyond}
Boyer,~M.~A.; Marsalek,~O.; Heindel,~J.~P.; Markland,~T.~E.; McCoy,~A.~B.; Xantheas,~S.~S. Beyond Badger’s rule: The origins and generality of the structure--spectra relationship of aqueous hydrogen bonds. \emph{The Journal of Physical Chemistry Letters} \textbf{2019}, \emph{10}, 918--924\relax
\mciteBstWouldAddEndPuncttrue
\mciteSetBstMidEndSepPunct{\mcitedefaultmidpunct}
{\mcitedefaultendpunct}{\mcitedefaultseppunct}\relax
\EndOfBibitem
\bibitem[Rackers \latin{et~al.}(2021)Rackers, Silva, Wang, and Ponder]{rackers2021polarizable}
Rackers,~J.~A.; Silva,~R.~R.; Wang,~Z.; Ponder,~J.~W. Polarizable water potential derived from a model electron density. \emph{Journal of chemical theory and computation} \textbf{2021}, \emph{17}, 7056--7084\relax
\mciteBstWouldAddEndPuncttrue
\mciteSetBstMidEndSepPunct{\mcitedefaultmidpunct}
{\mcitedefaultendpunct}{\mcitedefaultseppunct}\relax
\EndOfBibitem
\bibitem[Tang and Toennies(1992)Tang, and Toennies]{tang1992damping}
Tang,~K.; Toennies,~J.~P. The damping function of the van der Waals attraction in the potential between rare gas atoms and metal surfaces. \emph{Surface science} \textbf{1992}, \emph{279}, L203--L206\relax
\mciteBstWouldAddEndPuncttrue
\mciteSetBstMidEndSepPunct{\mcitedefaultmidpunct}
{\mcitedefaultendpunct}{\mcitedefaultseppunct}\relax
\EndOfBibitem
\bibitem[Tang and Toennies(1984)Tang, and Toennies]{tang1984improved}
Tang,~K.; Toennies,~J.~P. An improved simple model for the van der Waals potential based on universal damping functions for the dispersion coefficients. \emph{The Journal of chemical physics} \textbf{1984}, \emph{80}, 3726--3741\relax
\mciteBstWouldAddEndPuncttrue
\mciteSetBstMidEndSepPunct{\mcitedefaultmidpunct}
{\mcitedefaultendpunct}{\mcitedefaultseppunct}\relax
\EndOfBibitem
\bibitem[Demerdash \latin{et~al.}(2014)Demerdash, Yap, and Head-Gordon]{Demerdash2014}
Demerdash,~O.; Yap,~E.~H.; Head-Gordon,~T. Advanced potential energy surfaces for condensed phase simulation. \emph{Annu Rev Phys Chem} \textbf{2014}, \emph{65}, 149--74\relax
\mciteBstWouldAddEndPuncttrue
\mciteSetBstMidEndSepPunct{\mcitedefaultmidpunct}
{\mcitedefaultendpunct}{\mcitedefaultseppunct}\relax
\EndOfBibitem
\bibitem[Demerdash \latin{et~al.}(2018)Demerdash, Wang, and Head-Gordon]{Demerdash2018}
Demerdash,~O.; Wang,~L.-P.; Head-Gordon,~T. Advanced models for water simulations. \emph{WIREs Computational Molecular Science} \textbf{2018}, \emph{8}, e1355\relax
\mciteBstWouldAddEndPuncttrue
\mciteSetBstMidEndSepPunct{\mcitedefaultmidpunct}
{\mcitedefaultendpunct}{\mcitedefaultseppunct}\relax
\EndOfBibitem
\bibitem[Mortier \latin{et~al.}(1986)Mortier, Ghosh, and Shankar]{mortier1986electronegativity}
Mortier,~W.~J.; Ghosh,~S.~K.; Shankar,~S. Electronegativity-equalization method for the calculation of atomic charges in molecules. \emph{Journal of the American Chemical Society} \textbf{1986}, \emph{108}, 4315--4320\relax
\mciteBstWouldAddEndPuncttrue
\mciteSetBstMidEndSepPunct{\mcitedefaultmidpunct}
{\mcitedefaultendpunct}{\mcitedefaultseppunct}\relax
\EndOfBibitem
\bibitem[Mardirossian and Head-Gordon(2014)Mardirossian, and Head-Gordon]{Mardirossian2014}
Mardirossian,~N.; Head-Gordon,~M. Exploring the limit of accuracy for density functionals based on the generalized gradient approximation: Local, global hybrid, and range-separated hybrid functionals with and without dispersion corrections. \emph{J Chem Phys} \textbf{2014}, \emph{140}\relax
\mciteBstWouldAddEndPuncttrue
\mciteSetBstMidEndSepPunct{\mcitedefaultmidpunct}
{\mcitedefaultendpunct}{\mcitedefaultseppunct}\relax
\EndOfBibitem
\bibitem[Rappoport and Furche(2010)Rappoport, and Furche]{rappoport2010property}
Rappoport,~D.; Furche,~F. Property-optimized Gaussian basis sets for molecular response calculations. \emph{The Journal of chemical physics} \textbf{2010}, \emph{133}\relax
\mciteBstWouldAddEndPuncttrue
\mciteSetBstMidEndSepPunct{\mcitedefaultmidpunct}
{\mcitedefaultendpunct}{\mcitedefaultseppunct}\relax
\EndOfBibitem
\bibitem[Epifanovsky \latin{et~al.}(2021)Epifanovsky, Gilbert, Feng, Lee, Mao, Mardirossian, Pokhilko, White, Coons, Dempwolff, Gan, Hait, Horn, Jacobson, Kaliman, Kussmann, Lange, Lao, Levine, Liu, McKenzie, Morrison, Nanda, Plasser, Rehn, Vidal, You, Zhu, Alam, Albrecht, Aldossary, Alguire, Andersen, Athavale, Barton, Begam, Behn, Bellonzi, Bernard, Berquist, Burton, Carreras, Carter-Fenk, Chakraborty, Chien, Closser, Cofer-Shabica, Dasgupta, de~Wergifosse, Deng, Diedenhofen, Do, Ehlert, Fang, Fatehi, Feng, Friedhoff, Gayvert, Ge, Gidofalvi, Goldey, Gomes, González-Espinoza, Gulania, Gunina, Hanson-Heine, Harbach, Hauser, Herbst, Hernández~Vera, Hodecker, Holden, Houck, Huang, Hui, Huynh, Ivanov, Jasz, Ji, Jiang, Kaduk, Kähler, Khistyaev, Kim, Kis, Klunzinger, Koczor-Benda, Koh, Kosenkov, Koulias, Kowalczyk, Krauter, Kue, Kunitsa, Kus, Ladjánszki, Landau, Lawler, Lefrancois, Lehtola, \latin{et~al.} others]{Epifanovsky2021}
Epifanovsky,~E.; Gilbert,~A. T.~B.; Feng,~X.; Lee,~J.; Mao,~Y.; Mardirossian,~N.; Pokhilko,~P.; White,~A.~F.; Coons,~M.~P.; Dempwolff,~A.~L.; Gan,~Z.; Hait,~D.; Horn,~P.~R.; Jacobson,~L.~D.; Kaliman,~I.; Kussmann,~J.; Lange,~A.~W.; Lao,~K.~U.; Levine,~D.~S.; Liu,~J.; McKenzie,~S.~C.; Morrison,~A.~F.; Nanda,~K.~D.; Plasser,~F.; Rehn,~D.~R.; Vidal,~M.~L.; You,~Z.-Q.; Zhu,~Y.; Alam,~B.; Albrecht,~B.~J.; Aldossary,~A.; Alguire,~E.; Andersen,~J.~H.; Athavale,~V.; Barton,~D.; Begam,~K.; Behn,~A.; Bellonzi,~N.; Bernard,~Y.~A.; Berquist,~E.~J.; Burton,~H. G.~A.; Carreras,~A.; Carter-Fenk,~K.; Chakraborty,~R.; Chien,~A.~D.; Closser,~K.~D.; Cofer-Shabica,~V.; Dasgupta,~S.; de~Wergifosse,~M.; Deng,~J.; Diedenhofen,~M.; Do,~H.; Ehlert,~S.; Fang,~P.-T.; Fatehi,~S.; Feng,~Q.; Friedhoff,~T.; Gayvert,~J.; Ge,~Q.; Gidofalvi,~G.; Goldey,~M.; Gomes,~J.; González-Espinoza,~C.~E.; Gulania,~S.; Gunina,~A.~O.; Hanson-Heine,~M. W.~D.; Harbach,~P. H.~P.; Hauser,~A.; Herbst,~M.~F.; Hernández~Vera,~M.; Hodecker,~M.; Holden,~Z.~C.;
  Houck,~S.; Huang,~X.; Hui,~K.; Huynh,~B.~C.; Ivanov,~M.; Jasz,~A.; Ji,~H.; Jiang,~H.; Kaduk,~B.; Kähler,~S.; Khistyaev,~K.; Kim,~J.; Kis,~G.; Klunzinger,~P.; Koczor-Benda,~Z.; Koh,~J.~H.; Kosenkov,~D.; Koulias,~L.; Kowalczyk,~T.; Krauter,~C.~M.; Kue,~K.; Kunitsa,~A.; Kus,~T.; Ladjánszki,~I.; Landau,~A.; Lawler,~K.~V.; Lefrancois,~D.; Lehtola,~S.; others Software for the frontiers of quantum chemistry: An overview of developments in the Q-Chem 5 package. \emph{J. Chem. Phys.} \textbf{2021}, \emph{155}, 084801\relax
\mciteBstWouldAddEndPuncttrue
\mciteSetBstMidEndSepPunct{\mcitedefaultmidpunct}
{\mcitedefaultendpunct}{\mcitedefaultseppunct}\relax
\EndOfBibitem
\bibitem[Pracht \latin{et~al.}(2020)Pracht, Bohle, and Grimme]{pracht2020automated}
Pracht,~P.; Bohle,~F.; Grimme,~S. Automated exploration of the low-energy chemical space with fast quantum chemical methods. \emph{Physical Chemistry Chemical Physics} \textbf{2020}, \emph{22}, 7169--7192\relax
\mciteBstWouldAddEndPuncttrue
\mciteSetBstMidEndSepPunct{\mcitedefaultmidpunct}
{\mcitedefaultendpunct}{\mcitedefaultseppunct}\relax
\EndOfBibitem
\bibitem[Bannwarth \latin{et~al.}(2019)Bannwarth, Ehlert, and Grimme]{bannwarth2019gfn2}
Bannwarth,~C.; Ehlert,~S.; Grimme,~S. GFN2-xTB—An accurate and broadly parametrized self-consistent tight-binding quantum chemical method with multipole electrostatics and density-dependent dispersion contributions. \emph{Journal of chemical theory and computation} \textbf{2019}, \emph{15}, 1652--1671\relax
\mciteBstWouldAddEndPuncttrue
\mciteSetBstMidEndSepPunct{\mcitedefaultmidpunct}
{\mcitedefaultendpunct}{\mcitedefaultseppunct}\relax
\EndOfBibitem
\bibitem[Rakshit \latin{et~al.}(2019)Rakshit, Bandyopadhyay, Heindel, and Xantheas]{rakshit2019atlas}
Rakshit,~A.; Bandyopadhyay,~P.; Heindel,~J.~P.; Xantheas,~S.~S. Atlas of putative minima and low-lying energy networks of water clusters n= 3--25. \emph{The Journal of chemical physics} \textbf{2019}, \emph{151}\relax
\mciteBstWouldAddEndPuncttrue
\mciteSetBstMidEndSepPunct{\mcitedefaultmidpunct}
{\mcitedefaultendpunct}{\mcitedefaultseppunct}\relax
\EndOfBibitem
\bibitem[Bouchafra \latin{et~al.}(2018)Bouchafra, Shee, R{\'e}al, Vallet, and Severo Pereira~Gomes]{bouchafra2018predictive}
Bouchafra,~Y.; Shee,~A.; R{\'e}al,~F.; Vallet,~V.; Severo Pereira~Gomes,~A. Predictive simulations of ionization energies of solvated halide ions with relativistic embedded equation of motion coupled cluster theory. \emph{Physical review letters} \textbf{2018}, \emph{121}, 266001\relax
\mciteBstWouldAddEndPuncttrue
\mciteSetBstMidEndSepPunct{\mcitedefaultmidpunct}
{\mcitedefaultendpunct}{\mcitedefaultseppunct}\relax
\EndOfBibitem
\bibitem[Ren \latin{et~al.}(2011)Ren, Wu, and Ponder]{ren2011polarizable}
Ren,~P.; Wu,~C.; Ponder,~J.~W. Polarizable atomic multipole-based molecular mechanics for organic molecules. \emph{Journal of chemical theory and computation} \textbf{2011}, \emph{7}, 3143--3161\relax
\mciteBstWouldAddEndPuncttrue
\mciteSetBstMidEndSepPunct{\mcitedefaultmidpunct}
{\mcitedefaultendpunct}{\mcitedefaultseppunct}\relax
\EndOfBibitem
\bibitem[Chung \latin{et~al.}(2022)Chung, Wang, Rackers, and Ponder]{chung2022classical}
Chung,~M.~K.; Wang,~Z.; Rackers,~J.~A.; Ponder,~J.~W. Classical Exchange Polarization: An Anisotropic Variable Polarizability Model. \emph{The Journal of Physical Chemistry B} \textbf{2022}, \emph{126}, 7579--7594\relax
\mciteBstWouldAddEndPuncttrue
\mciteSetBstMidEndSepPunct{\mcitedefaultmidpunct}
{\mcitedefaultendpunct}{\mcitedefaultseppunct}\relax
\EndOfBibitem
\bibitem[Patkowski(2020)]{patkowski2020recent}
Patkowski,~K. Recent developments in symmetry-adapted perturbation theory. \emph{Wiley Interdisciplinary Reviews: Computational Molecular Science} \textbf{2020}, \emph{10}, e1452\relax
\mciteBstWouldAddEndPuncttrue
\mciteSetBstMidEndSepPunct{\mcitedefaultmidpunct}
{\mcitedefaultendpunct}{\mcitedefaultseppunct}\relax
\EndOfBibitem
\bibitem[Andrada and Foroutan-Nejad(2020)Andrada, and Foroutan-Nejad]{andrada2020energy}
Andrada,~D.~M.; Foroutan-Nejad,~C. Energy components in energy decomposition analysis (EDA) are path functions; why does it matter? \emph{Physical Chemistry Chemical Physics} \textbf{2020}, \emph{22}, 22459--22464\relax
\mciteBstWouldAddEndPuncttrue
\mciteSetBstMidEndSepPunct{\mcitedefaultmidpunct}
{\mcitedefaultendpunct}{\mcitedefaultseppunct}\relax
\EndOfBibitem
\bibitem[Buckingham(1979)]{buckingham1979polarizability}
Buckingham,~A.~D. Polarizability and hyperpolarizability. \emph{Philosophical Transactions of the Royal Society of London. Series A, Mathematical and Physical Sciences} \textbf{1979}, \emph{293}, 239--248\relax
\mciteBstWouldAddEndPuncttrue
\mciteSetBstMidEndSepPunct{\mcitedefaultmidpunct}
{\mcitedefaultendpunct}{\mcitedefaultseppunct}\relax
\EndOfBibitem
\bibitem[Wilson \latin{et~al.}(1996)Wilson, Madden, and Costa-Cabral]{wilson1996quadrupole}
Wilson,~M.; Madden,~P.~A.; Costa-Cabral,~B.~J. Quadrupole polarization in simulations of ionic systems: application to AgCl. \emph{the Journal of Physical Chemistry} \textbf{1996}, \emph{100}, 1227--1237\relax
\mciteBstWouldAddEndPuncttrue
\mciteSetBstMidEndSepPunct{\mcitedefaultmidpunct}
{\mcitedefaultendpunct}{\mcitedefaultseppunct}\relax
\EndOfBibitem
\bibitem[LaCour \latin{et~al.}(2023)LaCour, Heindel, and Head-Gordon]{lacour2023predicting}
LaCour,~R.~A.; Heindel,~J.~P.; Head-Gordon,~T. Predicting the Raman Spectra of Liquid Water with a Monomer-Field Model. \emph{The Journal of Physical Chemistry Letters} \textbf{2023}, \emph{14}, 11742--11749\relax
\mciteBstWouldAddEndPuncttrue
\mciteSetBstMidEndSepPunct{\mcitedefaultmidpunct}
{\mcitedefaultendpunct}{\mcitedefaultseppunct}\relax
\EndOfBibitem
\bibitem[Riera \latin{et~al.}(2017)Riera, Mardirossian, Bajaj, G{\"o}tz, and Paesani]{riera2017toward}
Riera,~M.; Mardirossian,~N.; Bajaj,~P.; G{\"o}tz,~A.~W.; Paesani,~F. Toward chemical accuracy in the description of ion--water interactions through many-body representations. Alkali-water dimer potential energy surfaces. \emph{The Journal of chemical physics} \textbf{2017}, \emph{147}\relax
\mciteBstWouldAddEndPuncttrue
\mciteSetBstMidEndSepPunct{\mcitedefaultmidpunct}
{\mcitedefaultendpunct}{\mcitedefaultseppunct}\relax
\EndOfBibitem
\bibitem[Caruso and Paesani(2021)Caruso, and Paesani]{caruso2021data}
Caruso,~A.; Paesani,~F. Data-driven many-body models enable a quantitative description of chloride hydration from clusters to bulk. \emph{The Journal of Chemical Physics} \textbf{2021}, \emph{155}\relax
\mciteBstWouldAddEndPuncttrue
\mciteSetBstMidEndSepPunct{\mcitedefaultmidpunct}
{\mcitedefaultendpunct}{\mcitedefaultseppunct}\relax
\EndOfBibitem
\bibitem[Caruso \latin{et~al.}(2022)Caruso, Zhu, Fulton, and Paesani]{caruso2022accurate}
Caruso,~A.; Zhu,~X.; Fulton,~J.~L.; Paesani,~F. Accurate modeling of bromide and iodide hydration with data-driven many-body potentials. \emph{The Journal of Physical Chemistry B} \textbf{2022}, \emph{126}, 8266--8278\relax
\mciteBstWouldAddEndPuncttrue
\mciteSetBstMidEndSepPunct{\mcitedefaultmidpunct}
{\mcitedefaultendpunct}{\mcitedefaultseppunct}\relax
\EndOfBibitem
\bibitem[Ponder \latin{et~al.}(2010)Ponder, Wu, Ren, Pande, Chodera, Schnieders, Haque, Mobley, Lambrecht, DiStasio~Jr, \latin{et~al.} others]{ponder2010current}
Ponder,~J.~W.; Wu,~C.; Ren,~P.; Pande,~V.~S.; Chodera,~J.~D.; Schnieders,~M.~J.; Haque,~I.; Mobley,~D.~L.; Lambrecht,~D.~S.; DiStasio~Jr,~R.~A.; others Current status of the AMOEBA polarizable force field. \emph{The journal of physical chemistry B} \textbf{2010}, \emph{114}, 2549--2564\relax
\mciteBstWouldAddEndPuncttrue
\mciteSetBstMidEndSepPunct{\mcitedefaultmidpunct}
{\mcitedefaultendpunct}{\mcitedefaultseppunct}\relax
\EndOfBibitem
\bibitem[Zhang \latin{et~al.}(2018)Zhang, Lu, Jing, Wu, Piquemal, Ponder, and Ren]{zhang2018amoeba}
Zhang,~C.; Lu,~C.; Jing,~Z.; Wu,~C.; Piquemal,~J.-P.; Ponder,~J.~W.; Ren,~P. AMOEBA polarizable atomic multipole force field for nucleic acids. \emph{Journal of chemical theory and computation} \textbf{2018}, \emph{14}, 2084--2108\relax
\mciteBstWouldAddEndPuncttrue
\mciteSetBstMidEndSepPunct{\mcitedefaultmidpunct}
{\mcitedefaultendpunct}{\mcitedefaultseppunct}\relax
\EndOfBibitem
\bibitem[Werner \latin{et~al.}(2011)Werner, Knizia, and Manby]{werner2011explicitly}
Werner,~H.-J.; Knizia,~G.; Manby,~F.~R. Explicitly correlated coupled cluster methods with pair-specific geminals. \emph{Molecular Physics} \textbf{2011}, \emph{109}, 407--417\relax
\mciteBstWouldAddEndPuncttrue
\mciteSetBstMidEndSepPunct{\mcitedefaultmidpunct}
{\mcitedefaultendpunct}{\mcitedefaultseppunct}\relax
\EndOfBibitem
\bibitem[Mahan(1980)]{mahan1980polarizability}
Mahan,~G. Polarizability of ions in crystals. \emph{Solid State Ionics} \textbf{1980}, \emph{1}, 29--45\relax
\mciteBstWouldAddEndPuncttrue
\mciteSetBstMidEndSepPunct{\mcitedefaultmidpunct}
{\mcitedefaultendpunct}{\mcitedefaultseppunct}\relax
\EndOfBibitem
\bibitem[Jacucci \latin{et~al.}(1976)Jacucci, McDonald, and Rahman]{jacucci1976effects}
Jacucci,~G.; McDonald,~I.; Rahman,~A. Effects of polarization on equilibrium and dynamic properties of ionic systems. \emph{Physical Review A} \textbf{1976}, \emph{13}, 1581\relax
\mciteBstWouldAddEndPuncttrue
\mciteSetBstMidEndSepPunct{\mcitedefaultmidpunct}
{\mcitedefaultendpunct}{\mcitedefaultseppunct}\relax
\EndOfBibitem
\bibitem[Pyper \latin{et~al.}(1992)Pyper, Pike, and Edwards]{pyper1992polarizabilities}
Pyper,~N.; Pike,~C.; Edwards,~P. The polarizabilities of species present in ionic solutions. \emph{Molecular Physics} \textbf{1992}, \emph{76}, 353--372\relax
\mciteBstWouldAddEndPuncttrue
\mciteSetBstMidEndSepPunct{\mcitedefaultmidpunct}
{\mcitedefaultendpunct}{\mcitedefaultseppunct}\relax
\EndOfBibitem
\bibitem[Li \latin{et~al.}(2017)Li, Zhuang, Lu, Wang, and An]{li2017accurate}
Li,~M.; Zhuang,~B.; Lu,~Y.; Wang,~Z.-G.; An,~L. Accurate determination of ion polarizabilities in aqueous solutions. \emph{The Journal of Physical Chemistry B} \textbf{2017}, \emph{121}, 6416--6424\relax
\mciteBstWouldAddEndPuncttrue
\mciteSetBstMidEndSepPunct{\mcitedefaultmidpunct}
{\mcitedefaultendpunct}{\mcitedefaultseppunct}\relax
\EndOfBibitem
\end{mcitethebibliography}
\end{document}


\title{Supplementary Information:\\ Completely Multipolar Model for Many-Body Water-Ion and Ion-Ion Interactions}
\author{Joseph Heindel$^{1,2}$, Lukas Kim$^{1}$, Martin Head-Gordon$^{1,2}$, Teresa Head-Gordon$^{1,2,3}$}
\date{\vspace{-10ex}}
\maketitle
\noindent
\begin{center}
$^1$Kenneth S. Pitzer Theory Center and Department of Chemistry\\
$^2$Chemical Sciences Division, Lawrence Berkeley National Laboratory\\
$^3$Departments of Bioengineering and Chemical and Biomolecular Engineering\\
University of California, Berkeley, CA, USA

corresponding author: thg@berkeley.edu
\end{center}

\vspace{10mm}
\begin{table}
    \centering
    \begin{tabular}{lccc}
          \multicolumn{4}{c}{RMSDs of \ce{Z^{+/-}(H_2O)_{5-16}} between FFs and DFT} \\\hline
         Ion & MBX & CMM & AMOEBA \\\hline
         \ce{Li^+} & 0.0525/0.0768 & 0.0738/0.0962 & 0.0919/0.1122 \\
         \ce{Na^+} & 0.0535/0.1031 & 0.0773/0.1182 & 0.0843/0.1215 \\
         \ce{K^+} & 0.0614/0.1258 & 0.0780/0.1326 & 0.0763/0.1135 \\
         \ce{Rb^+} & 0.0818/0.1813 & 0.0812/0.1472 & 0.0985/0.1666 \\
         \ce{Cs^+} & 0.0615/0.1142 & 0.0683/0.1192 & 0.0762/0.1419 \\\hline
         \ce{F^-} & 0.0341/0.0525 & 0.0758/0.1034 & 0.0799/0.0966 \\
         \ce{Cl^-} & 0.0420/0.0552 & 0.0570/0.0631 & 0.0683/0.0746 \\
         \ce{Br^-} & 0.0348/0.0460 & 0.0659/0.0743 & 0.0779/0.0917 \\
         \ce{I^-} & 0.0527/0.0711 & 0.0865/0.1026 & 0.0887/0.1047 \\\hline
    \end{tabular}
    \caption{Median/mean RMSD between $\omega$B97X-V/def2-TZVPPD clusters of \ce{Z^{+/-}(H_2O)_{5-16}}. }
    \label{tab:rmsds_ion_water}
\end{table}

\begin{figure}[H]
    \includegraphics*[width=\textwidth]{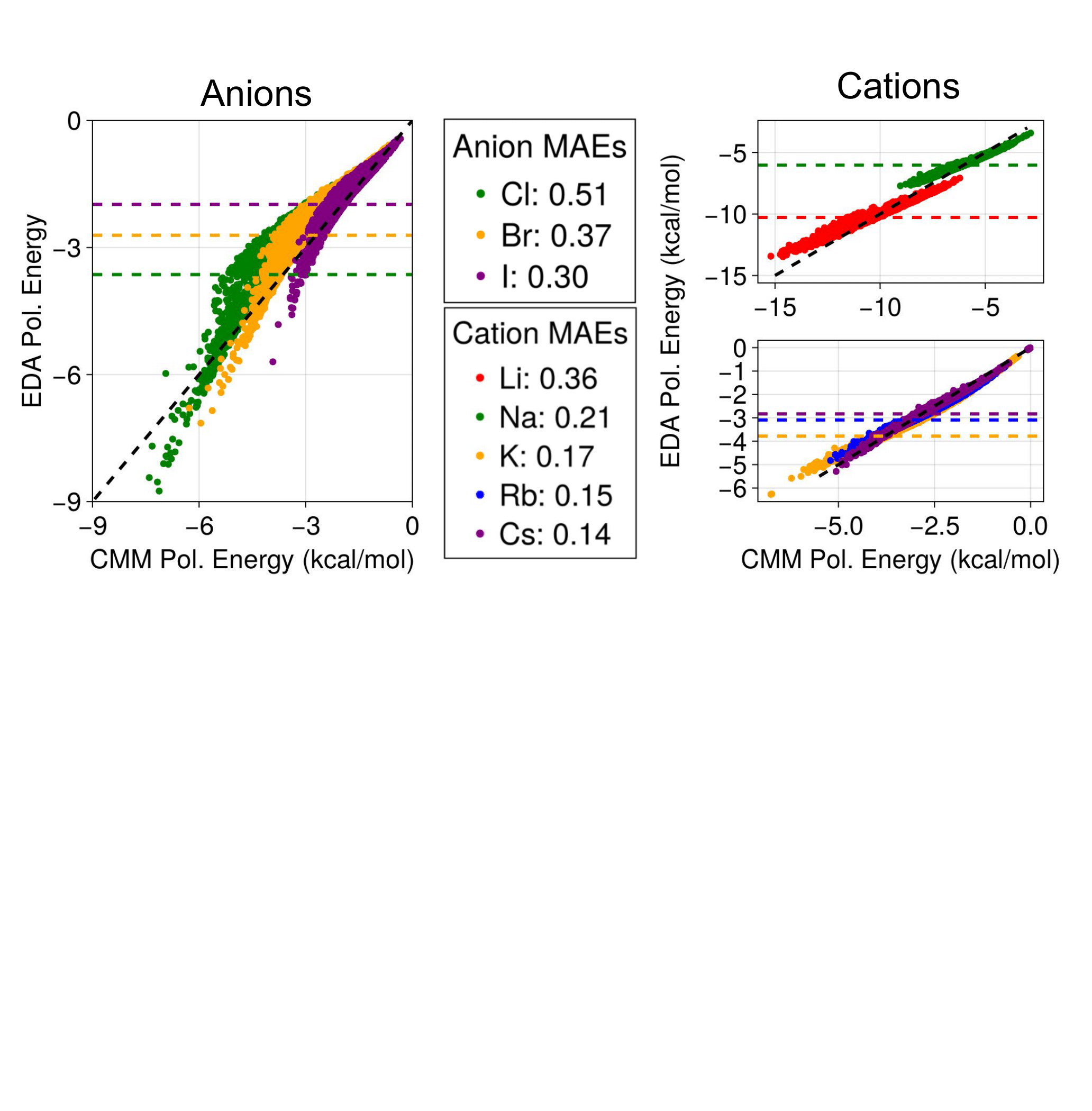}
    \caption{Difference in the polarizability parallel to the dimers of \ce{Li^+} and \ce{Cs^+} paired with either \ce{H_2O} or \ce{I^-}. On a percentage basis, the polarizability change between \ce{Li^+(H_2O)} is roughly the same as \ce{Cs^+I^-}. Clearly, the absolute change is much larger for the latter than the former. This perspective makes it clear that partial covalency has a much larger energetic effect for ion pairs despite the ratio of changes being rather similar.}
\end{figure}

\begin{figure}[H]
    \includegraphics*[width=\textwidth]{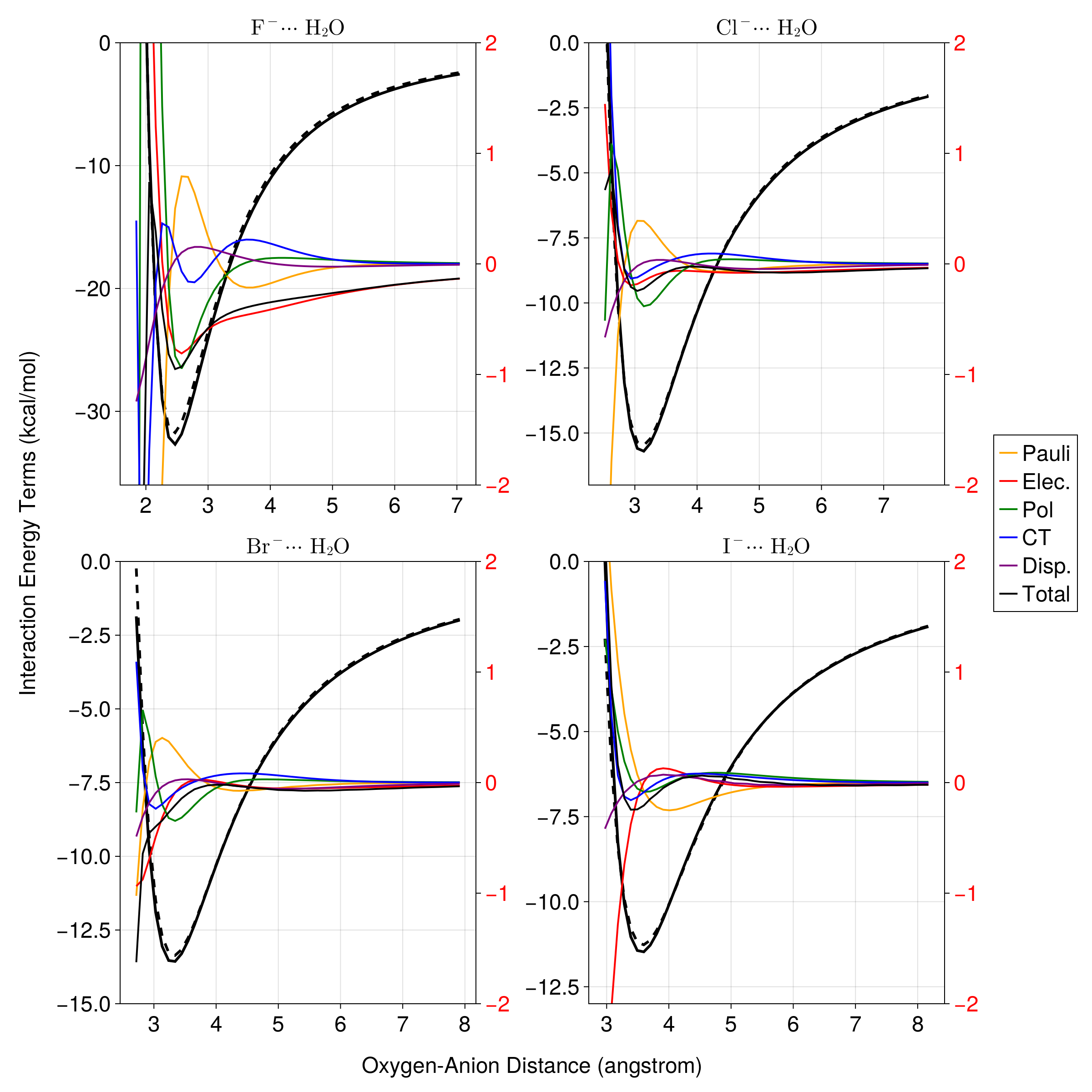}
    \caption{Scans of the total energy of halide-water dimers with the CMM model (solid black lines) compared to $\omega$B97X-V/def2-QZVPPD (dashed black line). The thin solid lines are the difference of each EDA term predicted by CMM from $\omega$B97X-V. The energy differences should be referenced to the axis on the right and the total energies to the axis on the left.}
\end{figure}

\begin{figure}[H]
    \includegraphics*[width=\textwidth]{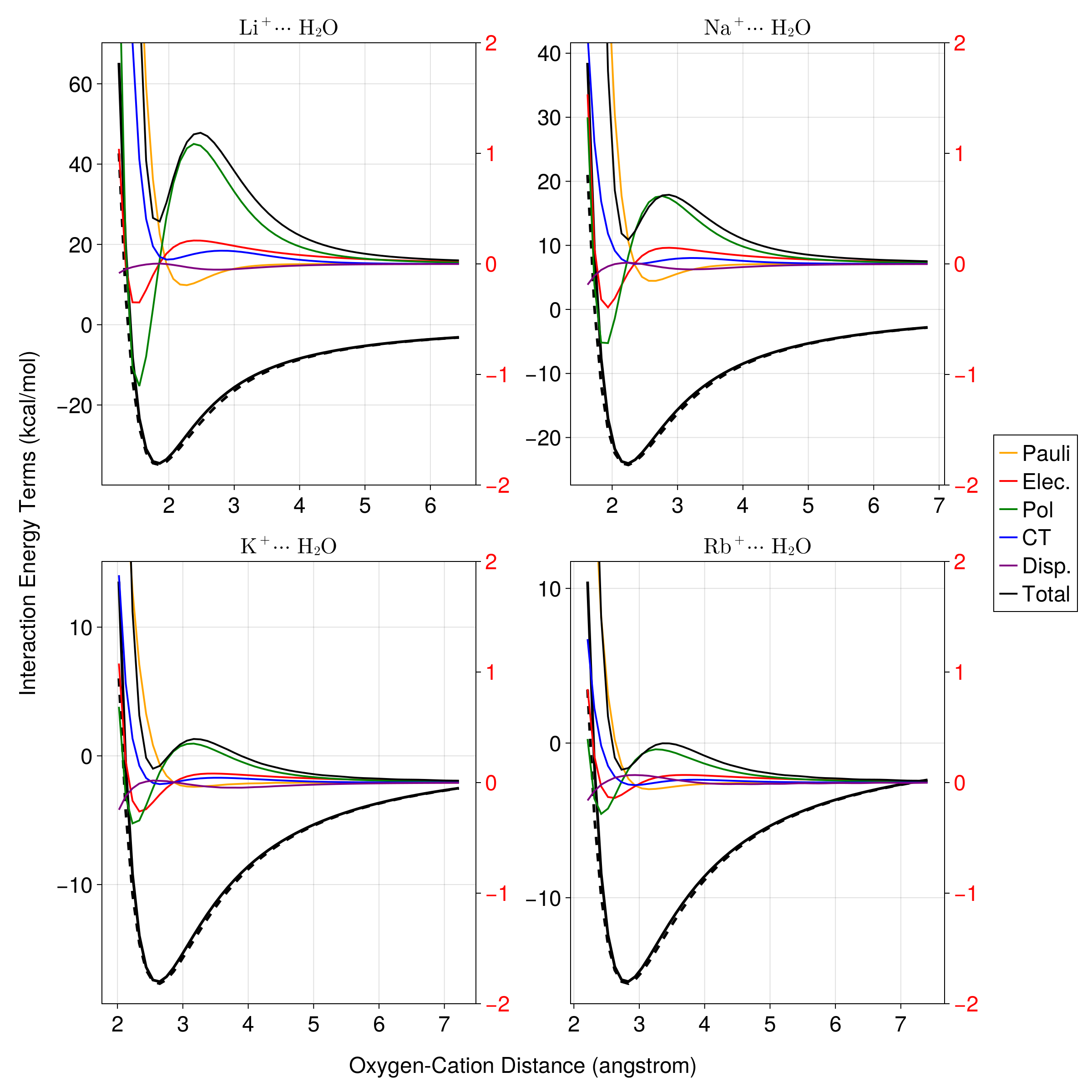}
    \caption{Scans of the total energy of alkali-water dimers with the CMM model (solid black lines) compared to $\omega$B97X-V/def2-QZVPPD (dashed black line). The thin solid lines are the difference of each EDA term predicted by CMM from $\omega$B97X-V. The energy differences should be referenced to the axis on the right and the total energies to the axis on the left.}
\end{figure}

\begin{figure}[H]
    \includegraphics*[width=\textwidth]{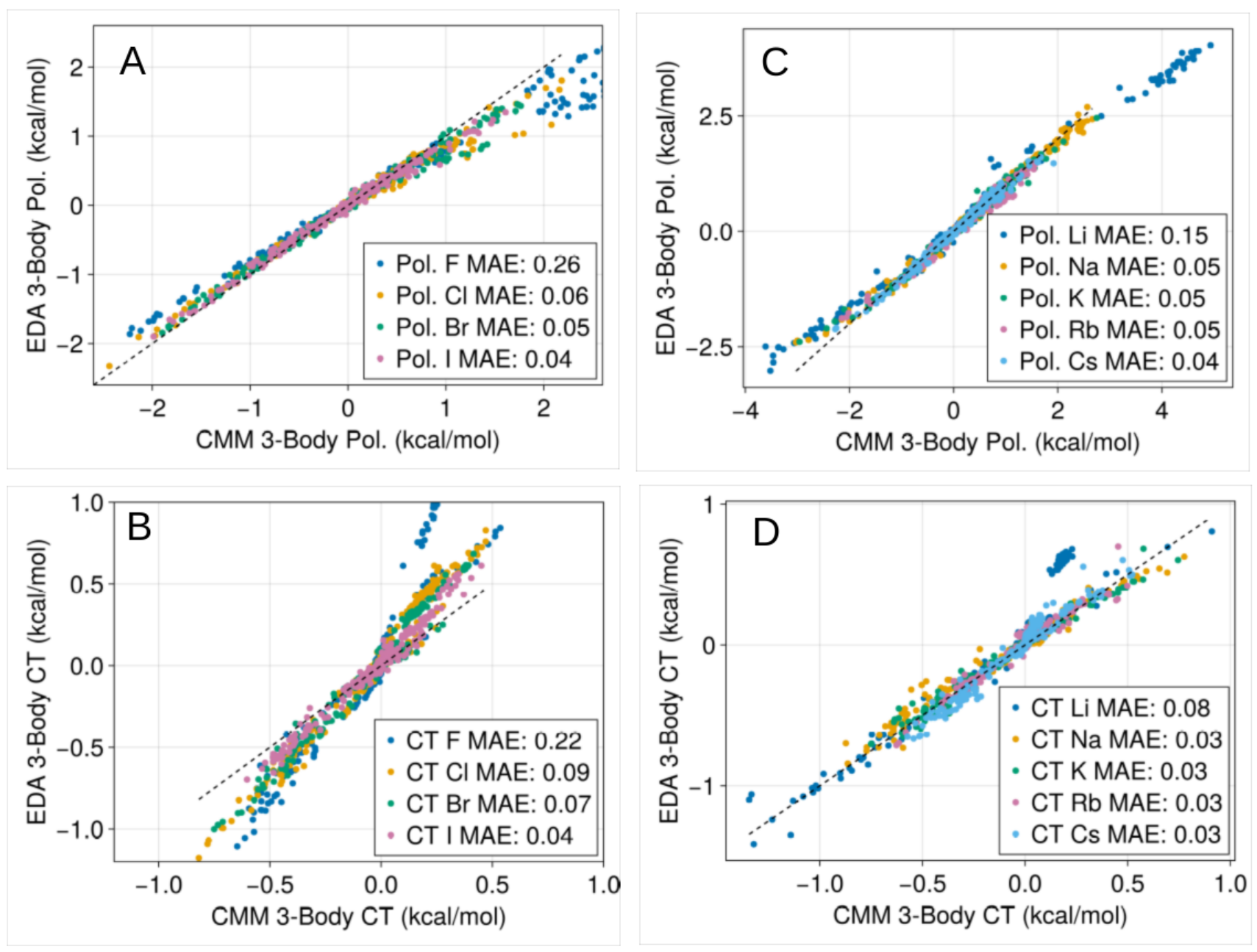}
    \caption{Correlation between three-body polarization and charge transfer energy from CMM and $\omega$B97X-V/def2-QZVPPD for anion-water and cation-water trimers taken from ion-water clusters. See main text for details. The legend shows the mean absolute error over all data points shown for each ion.}
    \label{fig:separated_three_body}
\end{figure}

\begin{figure}[H]
    \centering
    \includegraphics[width=1\linewidth]{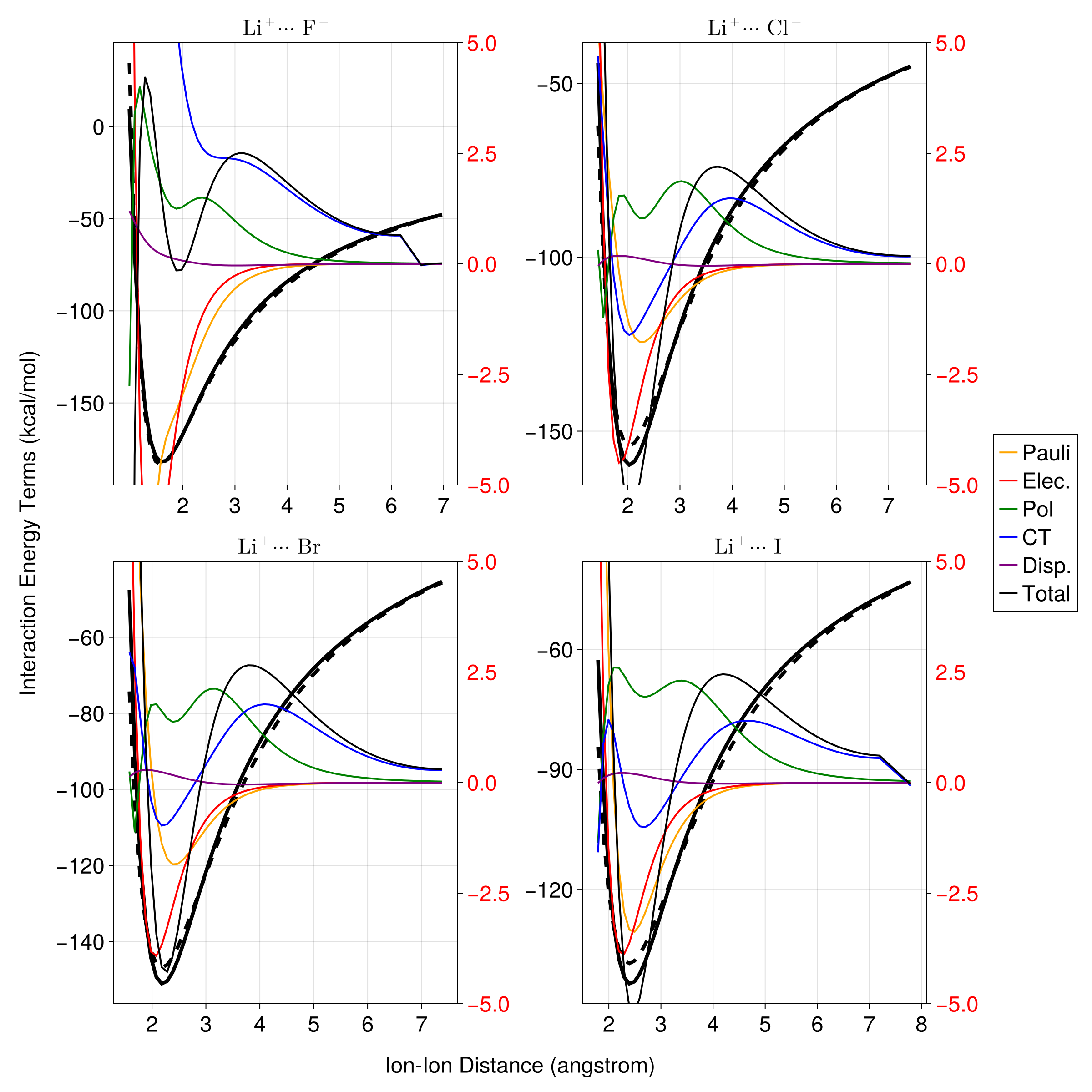}
    \caption{Scan of \ce{Li^+} paired with \ce{F^-}, \ce{Cl^-}, \ce{Br^-}, and \ce{I^-} using $\omega$B97X-V/def2-QZVPPD. The CMM total energy is shown with the thick solid black line while the total energy from DFT is dashed. The thin colored lines show the error in individual EDA terms predcted by CMM and should be referred to the axis on the right.}
\end{figure}

\begin{figure}[H]
    \centering
    \includegraphics[width=1\linewidth]{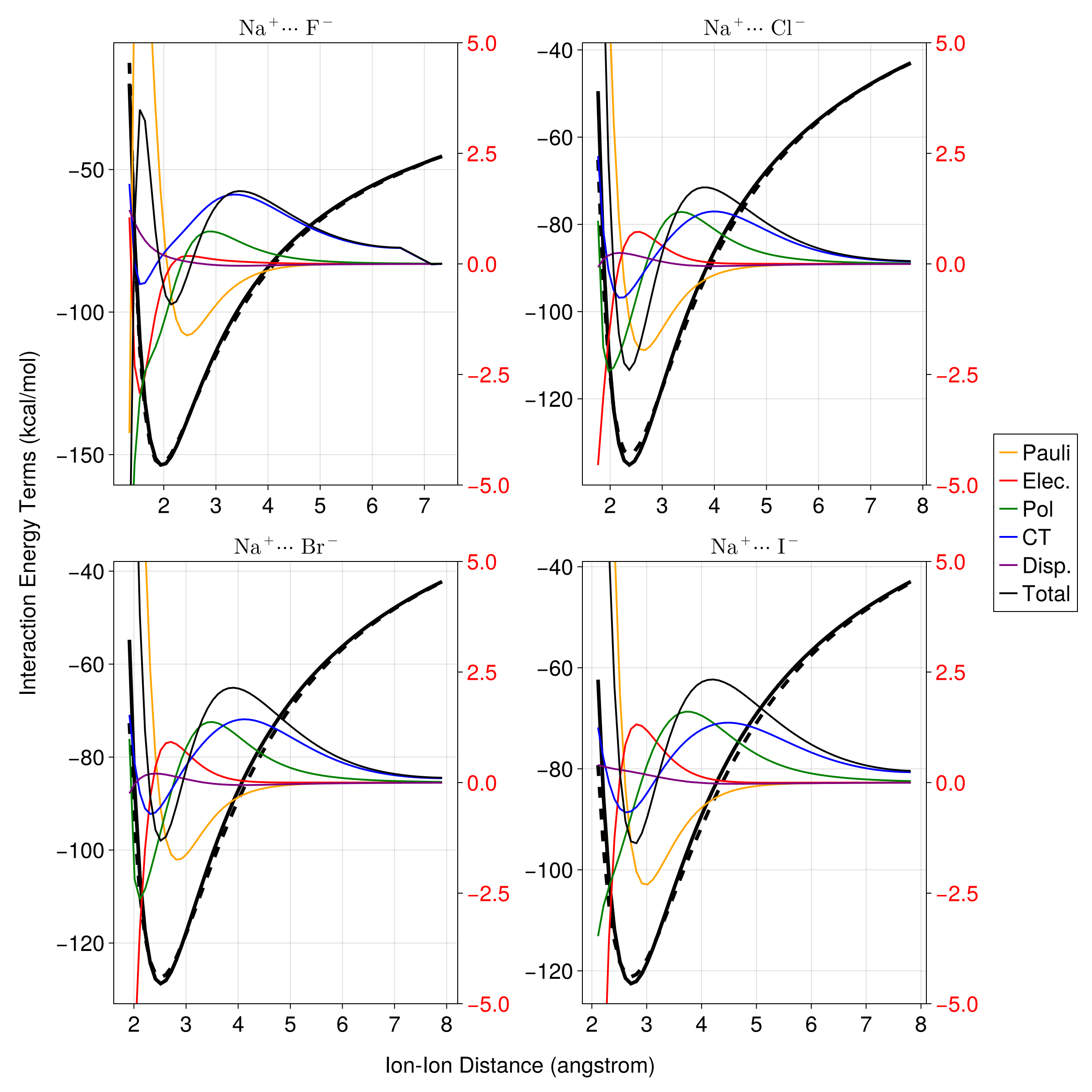}
    \caption{Scan of \ce{Na^+} paired with \ce{F^-}, \ce{Cl^-}, \ce{Br^-}, and \ce{I^-} using $\omega$B97X-V/def2-QZVPPD. The CMM total energy is shown with the thick solid black line while the total energy from DFT is dashed. The thin colored lines show the error in individual EDA terms predcted by CMM and should be referred to the axis on the right.}
\end{figure}

\begin{figure}[H]
    \centering
    \includegraphics[width=1\linewidth]{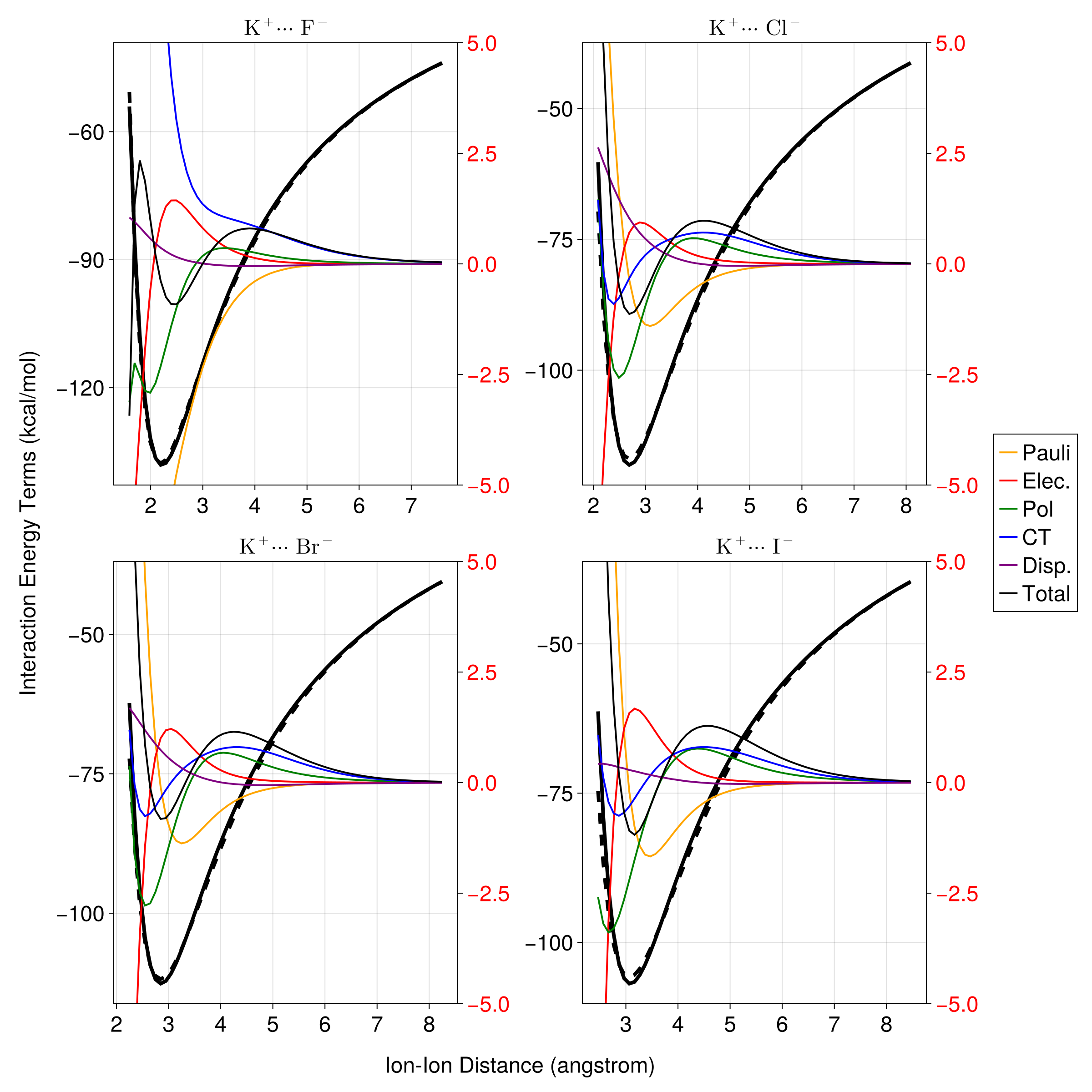}
    \caption{Scan of \ce{K^+} paired with \ce{F^-}, \ce{Cl^-}, \ce{Br^-}, and \ce{I^-} using $\omega$B97X-V/def2-QZVPPD. The CMM total energy is shown with the thick solid black line while the total energy from DFT is dashed. The thin colored lines show the error in individual EDA terms predcted by CMM and should be referred to the axis on the right.}
\end{figure}

\begin{figure}[H]
    \centering
    \includegraphics[width=1\linewidth]{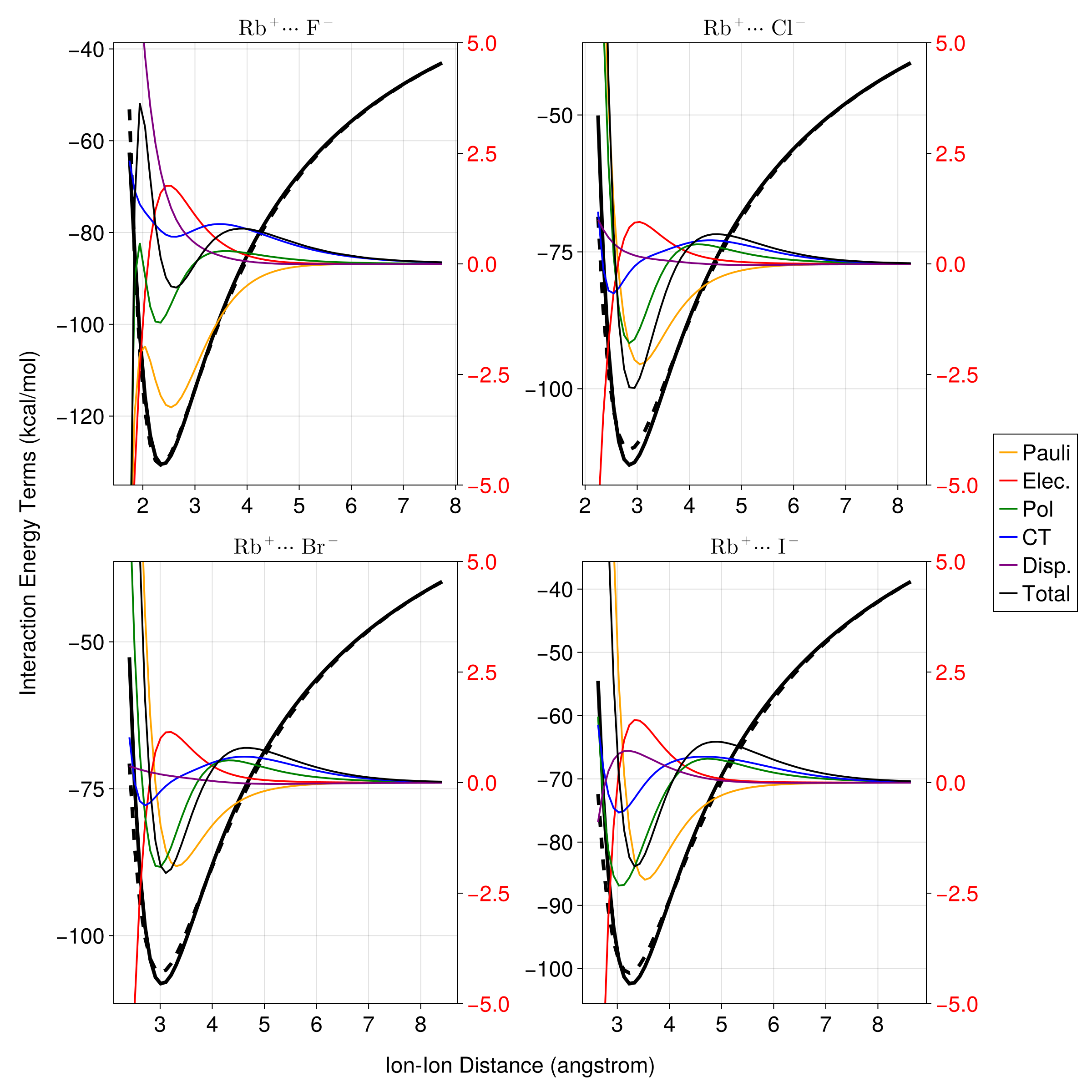}
    \caption{Scan of \ce{Rb^+} paired with \ce{F^-}, \ce{Cl^-}, \ce{Br^-}, and \ce{I^-} using $\omega$B97X-V/def2-QZVPPD. The CMM total energy is shown with the thick solid black line while the total energy from DFT is dashed. The thin colored lines show the error in individual EDA terms predcted by CMM and should be referred to the axis on the right.}
\end{figure}

\begin{figure}[H]
    \centering
    \includegraphics[width=1\linewidth]{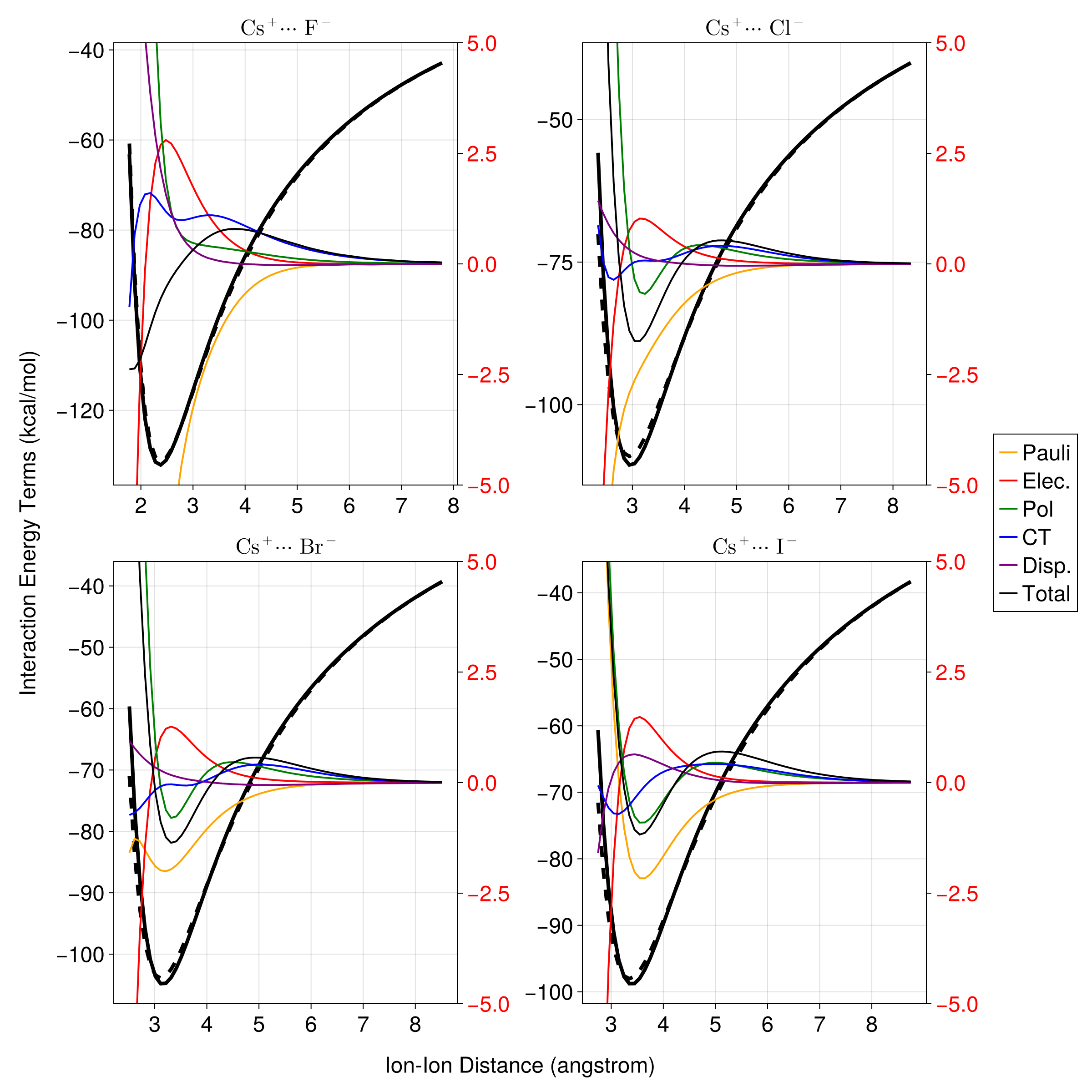}
    \caption{Scan of \ce{Cs^+} paired with \ce{F^-}, \ce{Cl^-}, \ce{Br^-}, and \ce{I^-} using $\omega$B97X-V/def2-QZVPPD. The CMM total energy is shown with the thick solid black line while the total energy from DFT is dashed. The thin colored lines show the error in individual EDA terms predcted by CMM and should be referred to the axis on the right.}
\end{figure}

\bibliographystyle{unsrtnat}
\bibliography{references}